\documentclass[reprint,amsmath,amssymb,aps,prl]{revtex4-2}
\usepackage{graphicx}
\usepackage{dcolumn}
\usepackage{bm}
\usepackage{color}
\usepackage{units}
\usepackage{wasysym}
\usepackage{MnSymbol}
\usepackage{comment}
\usepackage{times}
\usepackage{soul}
\usepackage{upgreek}
\usepackage[unicode=true,pdfusetitle,bookmarks=false,colorlinks=true,citecolor=blue,urlcolor=blue,linkcolor=blue]{hyperref}
\newcommand{\fakesection}[1]{
  \par\refstepcounter{section}
  \sectionmark{#1}
  \addcontentsline{toc}{section}{\protect\numberline{\thesection}#1}
}

\begin{document}

\title{Theory of Nonequilibrium Crystallization and the Phase Diagram of Active Brownian Spheres}
\author{Daniel Evans}
\author{Ahmad K. Omar}
\email{aomar@berkeley.edu}
\affiliation{Department of Materials Science and Engineering, University of California, Berkeley, California 94720, USA}
\affiliation{Materials Sciences Division, Lawrence Berkeley National Laboratory, Berkeley, California 94720, USA}

\begin{abstract}
The crystallization of hard spheres at equilibrium is perhaps the most familiar example of an entropically-driven phase transition.
In recent years, it has become clear that activity can dramatically alter this order-disorder transition in unexpected ways.
The theoretical description of active crystallization has remained elusive as the traditional thermodynamic arguments that shape our understanding of passive freezing are inapplicable to active systems.
Here, we develop a statistical mechanical description of the one-body density field and a nonconserved order parameter field that represents local crystalline order.
We develop equations of state, guided by computer simulations, describing the crystallinity field which result in shifting the order-disorder transition to higher packing fractions with increasing activity.
We then leverage our recent dynamical theory of coexistence to construct the full phase diagram of active Brownian spheres, quantitatively recapitulating both the solid-fluid and liquid-gas coexistence curves and the solid-liquid-gas triple point.
\end{abstract}

\maketitle

\textit{Introduction.--} From motile bacteria~\cite{Petroff2015} and starfish embryos exhibiting chiral motion~\cite{Tan2022} to light-activated colloids~\cite{Palacci2013LivingSurfers} and self-propelling liquid droplets~\cite{Kichatov2021CrystallizationEmulsion}, living and synthetic systems comprised of active matter are routinely observed to crystallize.
Activity has been shown to have a striking effect on crystalline phases, both theoretically~\cite{Caprini2023, Hermann2024ActiveTheory, Shi2023} and computationally~\cite{Bialke2012, Turci2021, Omar2021pd, Galliano2023}.
For example, the solid-fluid coexistence curve of monodisperse hard spheres~\cite{Alder1957, Hoover1968, Pusey1986, Pusey1989, Auer2001, Torquato2002, Pusey2009, Richard2018, Richard20182} is profoundly altered~\cite{Omar2021pd} with finite activity causing the solid phase density to rapidly increase to maximal fcc packing ${( \phi^{\rm solid} \approx \phi^{\rm CP} \equiv 0.74 )}$ from its equilibrium value ${( \phi^{\rm solid} \approx 0.545 )}$.
While the equilibrium crystallization of hard spheres is entropically-driven, an understanding of the driving forces behind active crystallization has remained elusive.
A theory of the active solid-fluid phase diagram would elucidate this driving force and allow for the design of stable active solids~\cite{Ferrante2013CollectiveCrystals, Baconnier2022SelectiveSolids, Xu2023AutonomousSolids, Chao2024} from microscopic considerations.

In this Letter, we develop a multi-scale theory of the crystallization of 3D active Brownian particles (ABPs).
While active crystals have been described through phase field crystal models~\cite{Menzel2013, Menzel2014, Ophaus2018, Ophaus2021, Holl2024Motility-inducedCrystallites}, where fluid and crystal phases have uniform and oscillatory one-body density fields, respectively, it can be challenging to relate the coefficients of these models to microscopic details out of equilibrium.
Here, we model crystal and fluid phases through a phase field approach where the density field is uniform in both phases and crystallinity is captured through a nonconserved order parameter.
We develop a statistical mechanical description of the dynamics of the density and crystallinity fields and identify the equations of state (EOSs)~\footnote{We note that we use EOS for the singular ``equation of state'' and EOSs for the plural ``equations of state''.} that govern the evolution of these fields.
We propose physics-guided empirical EOSs that capture the correct physical limits, finding that increasing activity shifts the order-disorder transition to higher packing fractions.
We then apply our recently developed dynamical theory of phase coexistence~\cite{Evans2024} to predict the solid-fluid phase diagram, finding it to be in strong agreement with that from simulation.

\textit{Theory of Solid-Fluid Coexistence.--}
Without loss of generality, we consider systems of $N$ particles in a volume $V$ at a fixed activity where macroscopic crystalline and fluid phases coexist with a planar interface normal to the $z$-direction.
This state of coexistence is described by two order parameter fields: a conserved number density $\rho(z)$ [or equivalently the volume fraction $\phi(z)$], subject to the constraint $\int_V d \mathbf{x} \rho(z) = N$, and a nonconserved and wholly unconstrained local dimensionless ``crystallinity'' field $\psi_N(z)$.
It is sometimes convenient to work with a crystallinity density field defined as $\psi_V(z) = \psi_N(z)\rho(z)$.
One scalar field ($\psi_N$ or $\psi_V$) is all that is needed to describe a cubic single crystalline phase, where, for example, the Steinhardt-Nelson-Ronchetti $q_{n}$ order parameter~\cite{Steinhardt1983} appropriately describes crystals with $n$-fold orientational symmetry between neighboring bonds~\footnote{Non-cubic crystals (that compete with cubic crystals) and polycrystals generally require additional nonconserved order parameters that quantify order in each unique spatial direction~\cite{Moelans2008AnEvolution} and the orientation of the polycrystals~\cite{Warren2003}, respectively.}.

The solid-fluid coexistence criteria in the passive limit can be found through straightforward thermodynamic arguments.
Minimizing the system free energy subject to the appropriate constraints on the global number density and system volume (see Ref.~\cite{Evans2024}) results in four criteria that can be used to find the four unknown values (two order parameters in each phase).
Using the free energy density of a homogeneous system, $f(\rho, \psi_V)$, and defining the bulk chemical potential $\mu_{\rho}^{\rm bulk} \equiv \partial f / \partial \rho$, bulk ``crystallinity potential'' ${\mu_{\psi}^{\rm bulk} \equiv \partial f / \partial \psi_V}$, and bulk pressure ${p^{\rm bulk} \equiv \mu_{\rho}^{\rm bulk} \rho + \mu_{\psi}^{\rm bulk} \psi_V - f}$, the criteria take the form:
\begin{subequations}
    \label{eq:thermo_criteria_pre}
    \begin{align}
        \label{eq:thermo_criteria_pre1}
        & \mu^{\rm bulk}_{\rho} \left( \rho^f, \psi_V^f \right) = \mu^{\rm bulk}_{\rho} \left( \rho^s, \psi_V^s \right) = \mu^{\rm coexist}_{\rho}, \\
        \label{eq:thermo_criteria_pre2}
        & \mu^{\rm bulk}_{\psi} \left( \rho^f, \psi_V^f \right) = 0, \\
        \label{eq:thermo_criteria_pre3}
        & \mu^{\rm bulk}_{\psi} \left( \rho^s, \psi_V^s \right) = 0, \\
        \label{eq:thermo_criteria_pre4}
        & p^{\rm bulk} \left( \rho^f, \psi_V^f \right) = p^{\rm bulk} \left( \rho^s, \psi_V^s \right) = p^{\rm coexist},
    \end{align}
\end{subequations}
where the $f$ and $s$ superscripts indicate fluid and solid phases.

Differentials of the pressure and chemical potentials are related through the equilibrium Gibbs-Duhem relation, ${d\mu_{\rho}^{\rm bulk} = \upsilon d p^{\rm bulk} - \psi_N d \mu_{\psi}^{\rm bulk}}$ where $\upsilon \equiv \rho^{-1}$.
Introducing the parameterization $\psi_V^*(\rho)$ satisfying $\mu_{\psi}^{\rm bulk}(\rho, \psi^*_V)=0$ ensures that two of the criteria [Eqs.~\eqref{eq:thermo_criteria_pre2} and \eqref{eq:thermo_criteria_pre3}] are automatically satisfied and the remaining criteria can be expressed solely in terms of $\mu_{\rho}^{\rm bulk}$ and $\psi_V^*(\rho)$ through a Maxwell construction:
\begin{subequations}
    \label{eq:thermo_criteria}
    \begin{align}
        & p^{\rm bulk} \left( \rho^f, \psi_V^* \right) = p^{\rm bulk} \left( \rho^s, \psi_V^* \right) = p^{\rm coexist}, \\
        \label{eq:thermo_criteria_int}
        & \int_{\upsilon^f}^{\upsilon^s} \left( p^{\rm bulk}(\rho, \psi_V^*) - p^{\rm coexist} \right) d \upsilon = 0,
    \end{align}
\end{subequations}
where, for notational convenience, we have ceased explicitly indicating the density dependence of $\psi^*_V$.
There may be multiple solution branches for $\psi_V^*$ (e.g.,~a crystalline branch and a fluid branch), however the branch chosen in Eq.~\eqref{eq:thermo_criteria} and the integration bounds in Eq.~\eqref{eq:thermo_criteria_int} should match the solid-fluid coexistence under consideration here.

The absence of a nonequilibrium variational principle necessitates a different perspective for obtaining the coexistence criteria for active systems. 
We proceed by recognizing that the order parameter fields still must be at a steady-state, $\dot{\rho} = - \partial_z J_{\rho} = 0$ and $\dot{\psi}_V = - \partial_z J_{\psi} + s_{\psi} = 0$ where $\partial_z \equiv \partial / \partial z$, $J_{\rho}$ and $J_{\psi}$ are the respective fluxes of $\rho$ and $\psi_V$ in the $z$-direction, and $s_{\psi}$ is the generation term of $\psi_V$.
We recently developed a coexistence framework~\cite{Evans2024} (building on Refs.~\cite{Aifantis1983rule, Solon2018, Omar2023b} which describe a single conserved field) that applies to these systems when the crystallinity flux is negligible, $|\partial_z J_{\psi}| << |s_{\psi}|$, and the number density flux and crystallinity generation are linear in their respective forces, $J_{\rho} = L f_{\rho}$ and $s_{\psi} = M f_{\psi}$ where $L$ and $M$ are positive transport coefficients.
The flux-driving force can contain terms of first and third-order in gradients of $\rho$ and $\psi_V$, $f_{\rho} = f_{\rho}^{(1)} + f_{\rho}^{(3)}$, whereas the generation-driving force can only contain bulk and second-order contributions, $f_{\psi} = f_{\psi}^{\rm bulk} + f_{\psi}^{(2)}$.
We expect these conditions to be \textit{approximately} met for active crystallization and will compare the resulting phase diagram with that obtained from the naive application of the equilibrium criteria.

Within our framework, the coexistence criteria emerge from the two static force balance conditions, $f_\rho = 0$ and $f_\psi=0$ (we thus only require the form of these forces within a multiplicative factor and the precise form of $L$ and $M$ are inconsequential).
As the force driving the crystallinity source must have even spatial parity, its leading order contribution in terms of spatial gradients is a bulk equation of state, $f_\psi^{\rm bulk}$. 
The absence of any spatial gradients deep into the coexisting phases ensures that $f_\psi^{\rm bulk} = 0$ in both phases. 
These criteria are analogous to the vanishing crystallinity potential in coexisting phases at equilibrium [Eqs.~\eqref{eq:thermo_criteria_pre2} and \eqref{eq:thermo_criteria_pre3}] and motivate defining a ``crystallinity pseudopotential'' $u_\psi =-f_\psi$ such that $u_\psi=\mu_\psi$ in equilibrium and Model A dynamics are recovered~\cite{Hohenberg1977}.
The two criteria can now be expressed as ${u_{\psi}^{\rm bulk} (\rho^f, \psi_V^f)=u_{\psi}^{\rm bulk} (\rho^s, \psi_V^s)= 0}$.

The last two coexistence criteria should be analogous to the equilibrium criteria of equality of chemical potential and pressure across phases.
Unlike $f_\psi$, $f_\rho$ is an odd function of space.
Its leading order term in the gradient expansion is thus not a bulk equation of state. 
We therefore seek to transform the zero flux-driving force steady-state condition into a uniform pseudopotential condition, $\mathcal{T}_{\rho \rho} f_{\rho} + \mathcal{T}_{\rho \psi} \psi_V \partial_z f_{\psi} = \partial_z u_{\rho}$ where $\mathcal{T}_{ij}$ is the $ij$ component of the transformation tensor and $u_{\rho}$ is the ``chemical pseudopotential'' ($u_{\rho} = u_{\rho}^{\rm bulk} + u_{\rho}^{(2)}$) which must be spatially uniform.
Crucially, the transformation tensor must be solved for using a set of coupled partial differential equations found from our definitions of $f_\rho$, $f_\psi$, and $u_\rho$. 
These equations may not admit a solution which suggests that, for the system under consideration, it may not be possible to cast the coexistence criteria in terms of bulk state functions.
In the event a solution is found, the uniformity of $u_{\rho}$ results in a third criterion: $u_{\rho}^{\rm bulk}(\rho^f, \psi_V^f) = u_{\rho}^{\rm bulk}(\rho^s, \psi_V^s) = u_{\rho}^{\rm coexist}$, mirroring equality of $p^{\rm bulk}$ across phases in equilibrium [Eq.~\eqref{eq:thermo_criteria_pre4}].

The final criterion results from an ansatz of a nonequilibrium Gibbs-Duhem relation, $d \mathcal{G} = \mathcal{E}_{\rho} d u_{\rho} + \mathcal{E}_{\psi} d u_{\psi}$ where $\boldsymbol{\mathcal{E}}$ is the ``Maxwell construction vector'' and $\mathcal{G}$ is a ``global quantity'' ($\mathcal{G} = \mathcal{G}^{\rm bulk} + \mathcal{G}^{(2)}$).
In equilibrium, $\mathcal{G} = \mu_{\rho}$ such that the equilibrium Gibbs-Duhem relation is recovered.
Determining $\boldsymbol{\mathcal{E}}$ and $\mathcal{G}$ again requires solving a system of coupled partial differential equations where a solution is not guaranteed.
When there is a solution, the fourth and final coexistence criterion follows from the spatial uniformity of $\mathcal{G}$: $\mathcal{G}^{\rm bulk}(\rho^f, \psi_V^f) = \mathcal{G}^{\rm bulk}(\rho^s, \psi_V^s) = \mathcal{G}^{\rm coexist}$, mirroring equality of $\mu_{\rho}^{\rm bulk}$ across phases in equilibrium [Eq.~\eqref{eq:thermo_criteria_pre1}].

The nonequilibrium coexistence criteria, just as in equilibrium, can be expressed solely in terms of $\psi_V^*(\rho)$ (defined to satisfy $u_{\psi}^{\rm bulk}(\rho, \psi^*_V)=0$) and $u^{\rm bulk}_{\rho}$ in a generalized Maxwell construction by invoking the nonequilibrium Gibbs-Duhem relation:
\begin{subequations}
    \label{eq:active_criteria}
    \begin{align}
        & u_{\rho}^{\rm bulk} \left( \rho^f, \psi_V^* \right) = u_{\rho}^{\rm bulk} \left( \rho^s, \psi_V^* \right) = u_{\rho}^{\rm coexist}, \\
        \label{eq:active_criteria_int}
        & \int_{\mathcal{E}_{\rho}^f}^{\mathcal{E}_{\rho}^s} \left( u_{\rho}^{\rm bulk}(\rho, \psi_V^*) - u_{\rho}^{\rm coexist} \right) d \mathcal{E}_{\rho} = 0.
    \end{align}
\end{subequations}

While it may be possible to rigorously find $u_{\rho}$ for many systems, in some of these systems a full solution for $\boldsymbol{\mathcal{E}}$ and $\mathcal{G}$ that satisfy the nonequilibrium Gibbs-Duhem relation~\cite{Evans2024, Chiu2024TheoryCoexistence} cannot be found.
When this is the case, the generalized Maxwell construction in Eq.~\eqref{eq:active_criteria_int} is no longer a rigorously established criterion.
However, these systems may still admit solutions to the \textit{interfacial} nonequilibrium Gibbs-Duhem relation, $d \mathcal{G}^{(2)} = \mathcal{E}^{\rm int}_{\rho} d u_{\rho}^{(2)} + \mathcal{E}^{\rm int}_{\psi} d u_{\psi}^{(2)}$ (i.e.,~ignoring bulk contributions), which allows \textit{approximate} forms of the final coexistence criterion to be developed.
A particularly useful form is to evaluate the generalized Maxwell construction in Eq.~\eqref{eq:active_criteria_int} using $\mathcal{E}_{\rho}=\mathcal{E}_{\rho}^{\rm int}$ which is only guaranteed to be zero if $\psi^*_V$ coincides with the relationship satisfied along the spatial coexistence profiles (which are unknown \textit{a priori}) of $\rho$ and $\psi_V$~\cite{Evans2024}.
We can thus approximate the criteria as:
\begin{subequations}
    \label{eq:active_criteria_approx}
    \begin{align}
        & u_{\rho}^{\rm bulk} \left( \rho^f, \psi_V^*\right) = u_{\rho}^{\rm bulk} \left( \rho^s, \psi_V^* \right) = u_{\rho}^{\rm coexist}, \\
        \label{eq:active_criteria_int_approx}
        & \int_{\mathcal{E}_{\rho}^{{\rm int}, f}}^{\mathcal{E}_{\rho}^{{\rm int}, s}} \left( u_{\rho}^{\rm bulk}(\rho, \psi_V^*) - u_{\rho}^{\rm coexist} \right) d \mathcal{E}_{\rho}^{\rm int} \approx 0.
    \end{align}
\end{subequations}

\textit{Theory of Active Crystallization.--}
We now look to predict the solid-fluid phase diagram of ABPs using the theory described in the previous Section, which requires expressions for $f_{\rho}$ and $f_{\psi}$. 
We consider $N$ athermal active Brownian spheres with overdamped translational and rotational dynamics. 
The orientation $\mathbf{q}_i$ of the $i^{th}$ particle satisfies $\dot{\mathbf{q}}_i = \boldsymbol{\Omega}_i \times \mathbf{q}_i$ where $\boldsymbol{\Omega}_i$ is a stochastic angular velocity with zero mean and variance $\langle \boldsymbol{\Omega}_i (t) \boldsymbol{\Omega}_j(t')\rangle = 2\delta_{ij}\delta(t-t')\mathbf{I} / \tau_R$ ($\mathbf{I}$ is the identity tensor and $\tau_R$ is the orientational relaxation time).
The position $\mathbf{r}_i$ of the $i^{th}$ particle satisfies $\dot{\mathbf{r}}_i =  U_0 \mathbf{q} + \mathbf{F}_i^C/\zeta$, where $U_0$ is the active self-propulsion speed, $\zeta$ is the translational drag coefficient, and $\mathbf{F}_i^C$ is a repulsive pairwise force which results in effective hard-sphere volume exclusion with diameter $D$~\cite{Omar2021pd}.
The state of the system is fully described by two dimensionless quantities: the volume fraction $\phi \equiv N \pi D^3 / 6 V$ and the ratio of the active runlength ($\ell_0 \equiv U_0 \tau_R$) to particle diameter $\ell_0 / D$~\cite{Omar2021pd}.

With these equations of motion, we derive statistical mechanical expressions for the dynamics of $\rho$ and $\psi_V$ to identify $f_{\rho}$ and $f_{\psi}$ using an Irving-Kirkwood procedure~\cite{Irving1950}.
We provide the full details of this derivation in the Supplementary Material (SM)~\footnote{See the SM for supplementary derivations and analytical EOSs.} and outline our approach in Appendix~\ref{secSI:derivation}.
We find $u_{\rho}$ to exist at all activities, identifying each pseudopotential as:
\begin{subequations}
    \label{eq:pseudopotentials}
    \begin{align}
        & u_{\rho} = p_C + p_{\rm act}^{\rm bulk} - \frac{\ell^2_0 \overline{U}}{20} \partial_z \left( \overline{U} \partial_z p_C^{\rm bulk}\right), \\
        & u_{\psi} = s_C - \frac{\ell^2_0}{24} \partial_z \left[ \overline{U} \partial_z\left(\overline{U} s_C^{\rm bulk} \right) \right],
    \end{align}
\end{subequations}
where $p_C = p_C^{\rm bulk} + p_C^{(2)}$ is the conservative interaction pressure (including Korteweg-like interfacial terms~\cite{Korteweg1904}), ${p_{\rm act}^{\rm bulk} \equiv \rho \zeta U_0 \overline{U} / 6}$ is the bulk active pressure~\cite{Takatori2014}, $\overline{U}$ is a dimensionless EOS capturing the decrease in effective self-propulsion due to interparticle interactions, and ${s_C = s_C^{\rm bulk} + s_C^{(2)}}$ is the generation of $\psi_V$ due to conservative interactions.

The nonequilibrium Gibbs-Duhem relation cannot be satisfied with the pseudopotentials in Eq.~\eqref{eq:pseudopotentials} and hence \textit{exact} solid-fluid coexistence criteria generally do not exist for ABPs.
However, criteria can be found in certain limits. 
For low activities ($\ell_0 / D \rightarrow 0$), the coexistence criteria are exactly the equilibrium criteria, as anticipated.
In the high-activity limit ($\ell_0 / D \rightarrow \infty$) the nonequilibrium Gibbs-Duhem relation does not have an exact solution, however the \textit{interfacial} nonequilibrium Gibbs-Duhem relation can be satisfied with $\mathcal{E}_{\rho}^{\rm int} = p_C^{\rm bulk}$ and $\mathcal{E}_{\psi}^{\rm int}=0$ with criteria of the form of Eq.~\eqref{eq:active_criteria_approx}.

\begin{figure}
	\centering
	\includegraphics[width=.475\textwidth]{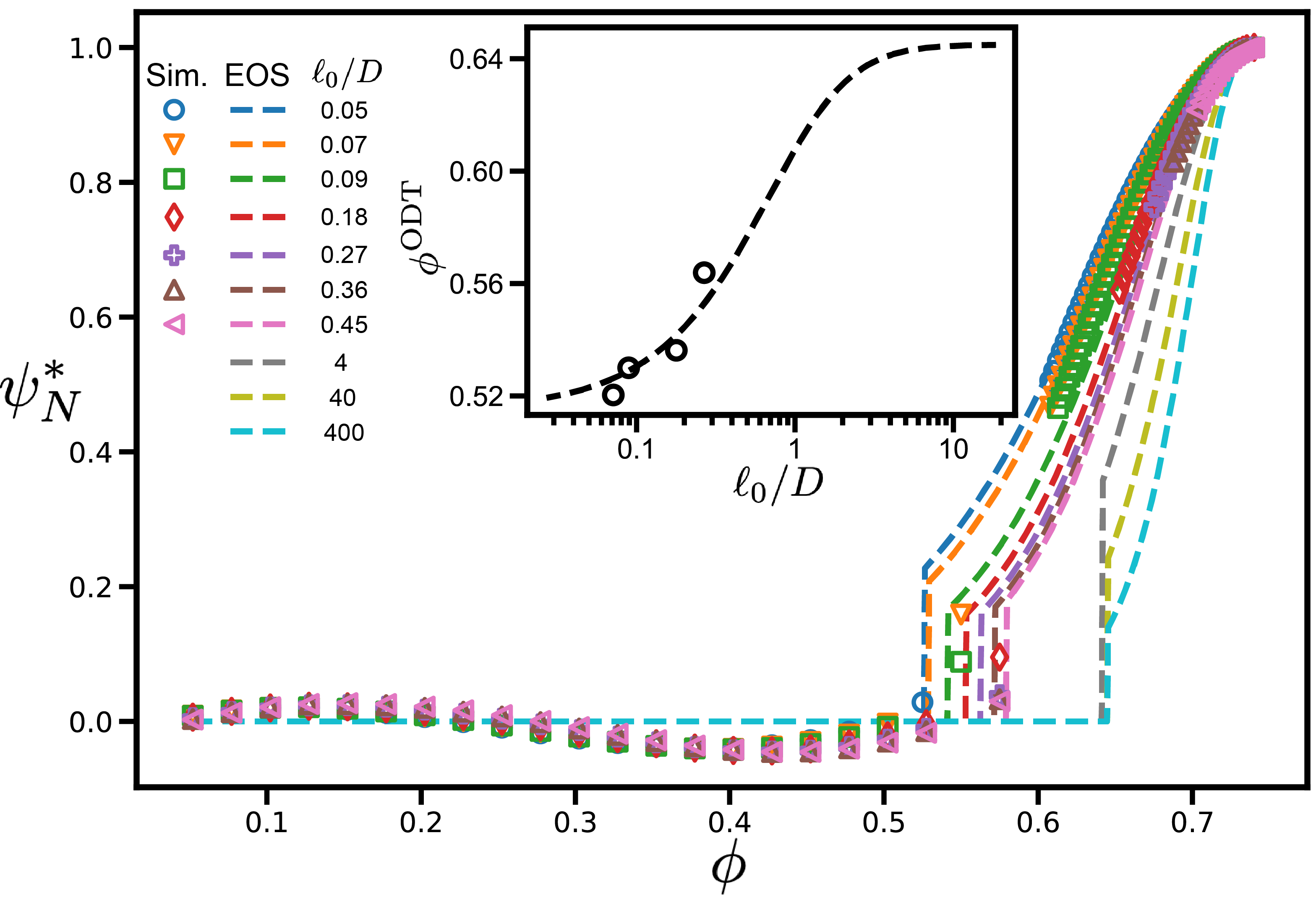}
	\caption{\protect\small{{Per-particle crystallinity, ${\psi_N^* ( \phi; \ell_0 / D )}$, of active hard spheres from Brownian dynamics simulation data (Sim.) and our EOS. The inset displays the accessible simulation data (symbols) and our EOS for ${\phi^{\rm ODT} ( \ell_0 / D )}$ (lines).}}}
	\label{fig:Psi}
\end{figure}

\begin{figure*}
	\centering
	\includegraphics[width=.95\textwidth]{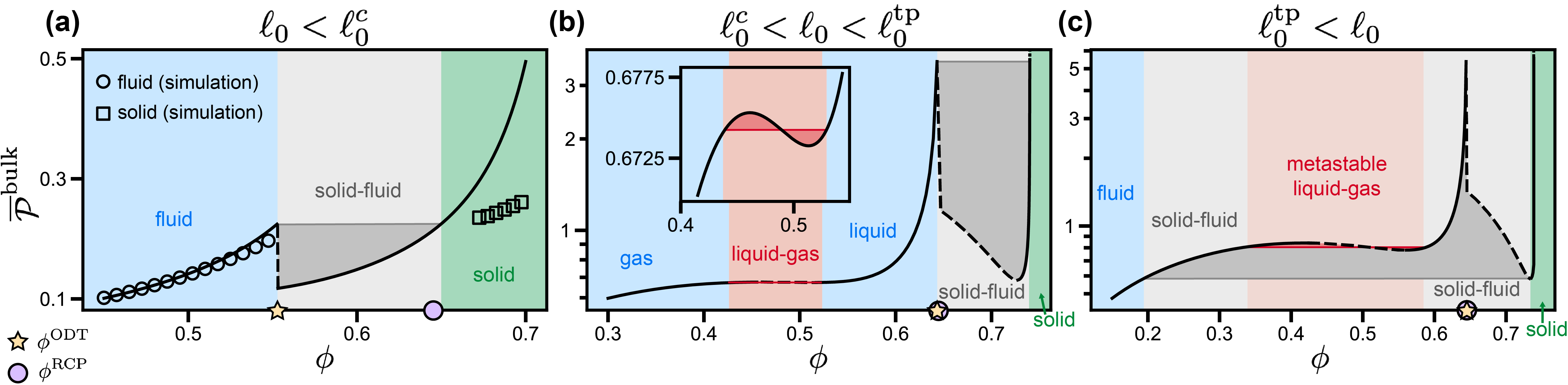}
	\caption{\protect\small{{Generalized Maxwell construction applied to the bulk dynamic pressure (nondimensionalized by ${6 \zeta U_0 / \pi D^2}$) of ABPs at three representative run lengths: (a) ${\ell_0 / D = 0.27}$, below the MIPS critical point (${\ell_0^c}$), (b) ${\ell_0 / D = 16.9}$, above ${\ell_0^c}$ but below the triple point (${\ell_0^{\rm tp}}$), and (c) ${\ell_0 / D = 22.3}$, above ${\ell_0^{\rm tp}}$. Dotted lines indicate the diverging pressure when the density of a solid is increased beyond close-packing. The red region in (c) is lighter than that in (b) as liquid-gas coexistence is metastable with respect to the globally stable solid-fluid coexistence in (c) whereas it is globally stable in (b).}}}
	\label{fig:Ps}
\end{figure*}

With the criteria established, we now require EOSs for ${p_C^{\rm bulk} ( \phi, \psi_V ; \ell_0 / D )}$, ${\overline{U} ( \phi, \psi_V ; \ell_0 / D )}$, and ${\psi^*_V ( \phi; \ell_0 / D )}$.
We determine ${\psi^*_V ( \phi; \ell_0 / D )}$ by computing the most probable crystallinity from 3D Brownian dynamics simulations~\cite{Anderson2020, Omar2021pd} of homogeneous systems (see the SM~\cite{Note3} for simulation details), defining ${\psi_N \equiv ( q_{12} - q_{12}^{\rm IG} ) / ( q_{12}^{\rm CP} - q_{12}^{\rm IG})}$ where ${q_{12}}$ is the per-particle Steinhardt-Nelson-Ronchetti order parameter~\cite{Steinhardt1983} and ${q_{12}^{\rm IG}}$ and ${q_{12}^{\rm CP}}$ are the values of ${q_{12}}$ in an ideal gas and close-packed fcc solid, respectively.
Figure~\ref{fig:Psi} displays ${\psi_N^*}$ obtained from simulation along with our fit.
For all activities, a disordered fluid (${\psi_N^*=0}$) and a perfectly ordered fcc crystal (${\psi_N^* =1}$) are found in the limits of ${\phi \rightarrow 0}$ and ${\phi \rightarrow \phi^{\rm CP}}$, respectively.
Furthermore, ${\psi_N^*}$ experiences a discontinuity at an activity-dependent volume fraction -- the \textit{order-disorder volume fraction}, ${\phi^{\rm ODT}}$ -- which must be less than or equal to random-close packing, ${\phi^{\rm RCP} \approx 0.645}$ (a fluid must begin to order when ${\phi > \phi^{\rm RCP}}$~\cite{Torquato2000}), and will ultimately lie within the solid-fluid binodal.
The activity dependence of ${\phi^{\rm ODT}}$ is thus crucial in determining ${\psi_N^*}$.
At low activities, ${\phi^{\rm ODT}}$ approaches the equilibrium hard sphere value of ${{0.515}}$.
With increasing activity, ${\phi^{\rm ODT}}$ monotonically increases before saturating at ${\phi^{\rm RCP}}$ at a remarkably low activity of ${{\ell_0 / D \approx 5}}$.
This activity-induced delay in the ordering transition is, as we will demonstrate, consistent with the reported dramatic shift of the solid-fluid binodal~\cite{Omar2021pd} upon departing from the reversible limit.

EOSs for ${p_C^{\rm bulk}}$ and ${\overline{U}}$ in a fluid of active Brownian spheres at activities ${\ell_0 / D \geq 1}$ and ${\psi_N=0}$ were recently developed~\cite{Omar2023b}. 
We extend these to nonzero ${\psi_N}$ and all ${\ell_0 / D}$ as detailed in the SM~\cite{Note3}.
For a fixed density and activity, increasing crystallinity results in additional free volume that \textit{increases} the active pressure while reducing the hard-sphere interaction pressure.
We ensure that in the limit of low activity, ${p_C^{\rm bulk}}$ recovers the equilibrium interaction pressure of hard spheres~\cite{Song1988}.
Figure~\ref{fig:Ps} shows the resulting EOS for the dynamic pressure, ${\mathcal{P}^{\rm bulk} \equiv p_C^{\rm bulk} + p_{\rm act}^{\rm bulk}}$ at $\psi^*_N(\phi)$, and the approximate generalized Maxwell construction [Eq.~\eqref{eq:active_criteria_approx}] in three distinct activity regimes.
The dashed line indicates the non-monotonic unstable region of the pressure.
At low activities, this occurs over an infinitesimally narrow region of volume fraction coinciding with ${\phi^{\rm ODT}}$ as the pressure experiences a discontinuity at this point~\cite{Note3} but is otherwise monotonically increasing with $\phi$ along $\psi^*_N$.
This discontinuity is consistent with the first-order nature of freezing/melting.
We emphasize that this ``pseudo''-spinodal does not imply that crystallization of a disordered fluid (${\psi_N=0}$) is a spontaneous process, but simply that homogeneous states at \textit{these values} of ${\phi}$ and ${\psi_N^*}$ are unstable. 

At an activity below the MIPS critical point (${\ell_0^c \approx 16.7 ~ D}$) solid-fluid coexistence is the only coexistence scenario, as shown in Fig.~\ref{fig:Ps}(a).
As $\mathcal{E}_{\rho}^{\rm int} = p_C^{\rm bulk}$ experiences a discontinuous downward jump at $\phi^{\rm ODT}$, the fluid density approaches $\phi^{\rm ODT}$ such that the integrand of Eq.~\eqref{eq:active_criteria_int_approx} remains finite at $\phi^{\rm ODT}$.
Above the critical point but below the triple point (${\ell_0^{\rm tp} \approx 18.3 ~ D}$), there are two distinct regions of coexistence [see Fig.~\ref{fig:Ps}(b)].
In this regime, the coexisting solid and liquid densities have shifted towards much higher volume fractions while MIPS occurs at lower volume fractions (below ${\phi^{\rm ODT}}$).
The two coexistence scenarios are separated by an appreciable gap in volume fractions.
As the activity is increased towards the triple point, the high density branch of the liquid-gas coexistence curve and the low density branch of the solid-fluid coexistence curve approach each other and coincide at the triple point. 
Above the triple point, the low density branch of the solid-fluid coexistence curve is now \textit{below} that of MIPS, with the former coexistence scenario engulfing the latter [see Fig.~\ref{fig:Ps}(c)].
Contrasting the maximization of the fluid density below the triple point, the fluid density is now minimized to offset the jump in $\mathcal{E}_{\rho}^{\rm int}$ at $\phi^{\rm ODT}$ in the generalized Maxwell construction.
Using simple arguments from large deviation theory, it was recently shown that solid-gas coexistence is stable over liquid-gas coexistence in this regime~\cite{Omar2021pd}.

\begin{figure}
	\centering
	\includegraphics[width=.475\textwidth]{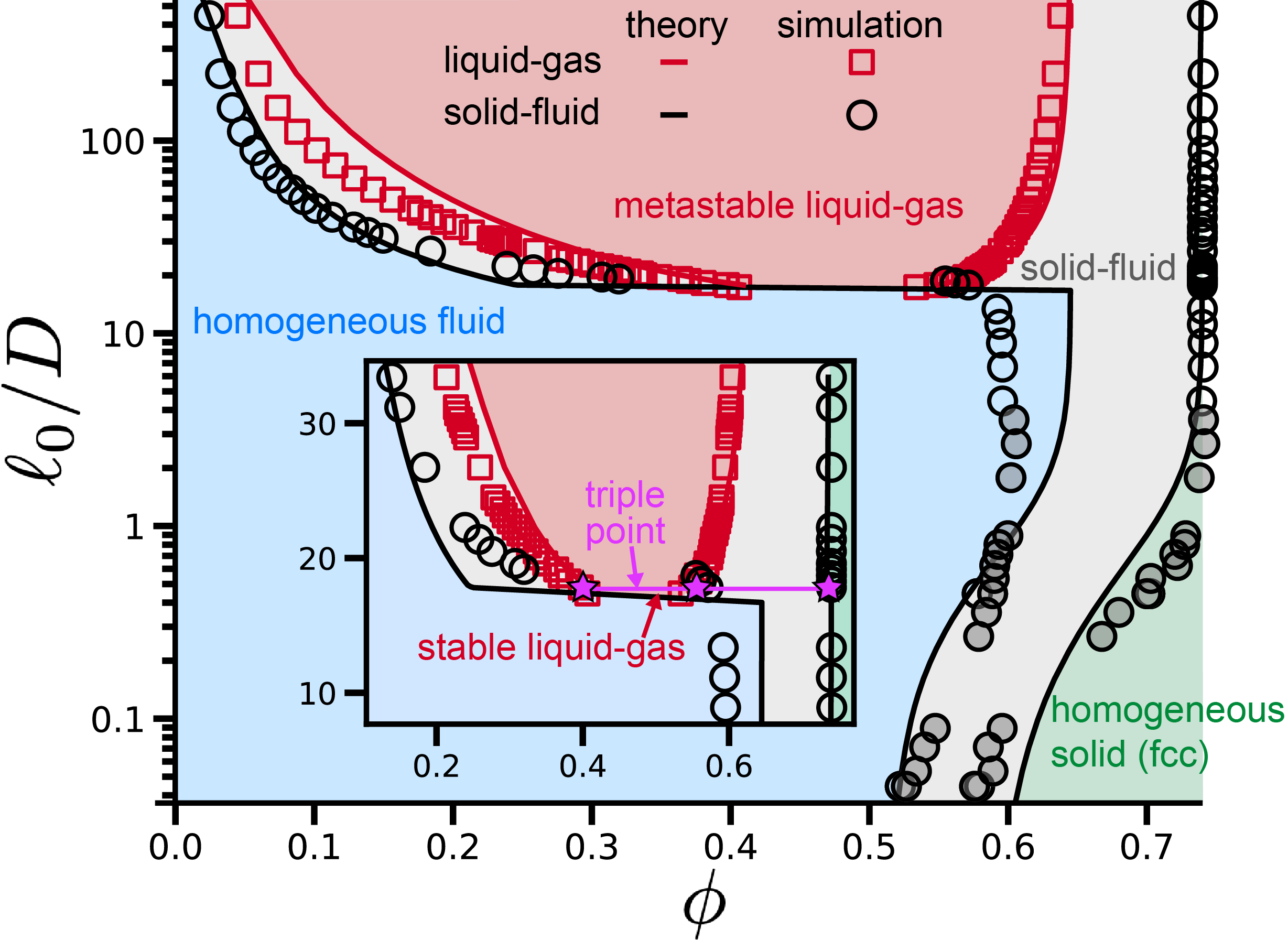}
	\caption{\protect\small{{Phase diagram of 3D ABPs including both solid-fluid and liquid-gas coexistence. Open circles are solid-fluid coexistence data from Ref.~\cite{Omar2021pd} while filled circles are data obtained in this study.}}}
	\label{fig:pd}
\end{figure}

Figure~\ref{fig:pd} shows the complete activity dependence of our predicted phase diagram [using Eq.~\eqref{eq:active_criteria_approx}] along with that obtained from computer simulations~\cite{Omar2021pd}.
In addition to naturally recovering the MIPS binodal, our theory nearly (especially with increasing activity) captures the solid-fluid binodal at all values of activity. 
While the predicted solid-fluid coexistence curve does not recover the passive hard sphere limit at vanishing run lengths, it does capture the rapid approach of the solid phase density towards close-packing at activities as low as ${\ell_0 / D \approx 1}$.
The theory correctly predicts the location of the solid-liquid-gas triple point and quantitatively recapitulates the solid-gas coexistence densities at high activities.
We note that the coexistence criteria used only become \textit{well-defined} in the high-activity limit, and are still approximate criteria in this limit.
Nevertheless, our coexistence theory is quantitatively accurate across a range of intermediate and high activities. 
To the best of our knowledge, our theory is the first to capture the binodals associated with both MIPS and active crystallization while making \textit{no appeals} to equilibrium thermodynamics.

Our nonequilibrium criteria predicts coexistence densities that are distinct from those naively predicted using the equilibrium criteria.
At low activities, both approaches result in similar predictions as shown in Appendix~\ref{secSI:comparecriteria}.
With increasing activity, continuing to erroneously use the equilibrium coexistence criteria is found to result in qualitative errors.
In fact, the equilibrium criteria predicts the solid to coexist with a liquid phase above the triple point, vastly overestimating the density of the fluid phase.
The equilibrium Gibbs-Duhem relation, and consequently the Maxwell equal-area construction, is thus violated at finite activity, making clear that use of the equilibrium Gibbs-Duhem relation to obtain active phase diagrams is formally incorrect and can result in significant error. 
As shown in Appendix~\ref{secSI:work}, the degree to which the equilibrium Gibbs-Duhem relation is violated can provide insight into the work required to move particles across the interface dividing two coexisting phases which is related to the polarization of particles along the interface.
Interestingly, our EOSs predict the polarization of the solid-fluid interface switches sign at an activity above the triple point, whereas the liquid-gas interface is polarized towards the liquid phase at all activities.

\textit{Discussion and Conclusions.--} We have applied a nonequilibrium coexistence theory to active crystallization, first deriving expressions for the evolution of the density and crystallinity fields and then developing physically and empirically motivated equations of state that capture the effects of activity on the order-disorder transition and crystalline order on the dynamic pressure.
We combine these equations of state with our coexistence criteria to quantitatively recapitulate the phase diagram of active Brownian spheres, demonstrating significant improvement over the binodals computed under the naive use of the equilibrium Maxwell construction.
We hope this work offers a concrete path towards the development of a general theory for nonequilibrium crystallization.

\begin{acknowledgements}
\textit{Acknowledgments.--}
D.E. acknowledges support from the U.S. Department of Defense through the National Defense Science and Engineering Graduate Fellowship Program.
This research used the Savio computational cluster resource provided by the Berkeley Research Computing program. 
\end{acknowledgements}

\section{End Matter}

\fakesection{Microscopic Derivation of Pseudopotentials} \label{secSI:derivation}
\textit{Appendix I: Microscopic Derivation of Pseudopotentials.--}
We now outline the derivation of $u_{\rho}$ and $u_{\psi}$ in Eq.~\ref{eq:pseudopotentials} beginning from the ABP equations of motion (a detailed derivation can be found in the SM~\cite{Note3}).
We do this following an Irving-Kirkwood procedure~\cite{Irving1950} in the same manner as was done to derive the density field dynamics of ABPs in Ref.~\cite{Omar2023b}, introducing the noise-averaged $N$-body distribution $P_N$ that satisfies the Fokker-Planck equation $\dot{P}_N = \mathcal{L} P_N$ where $\mathcal{L}$ is the Fokker-Planck operator associated with the ABP equations of motion.
For ABPs, the adjoint of the Fokker-Planck operator, $\mathcal{L}^{\dagger}$, is found to be~\cite{Omar2023b}:
\begin{equation}
    \mathcal{L}^{\dagger} \equiv \sum_i \left[  \left( U_0 \mathbf{q}_i + \frac{1}{\zeta} \mathbf{F}_i^C \right) \cdot \frac{\partial}{\partial \mathbf{r}_i} + \tau_R^{-1} \boldsymbol{\nabla}_i^R \cdot \boldsymbol{\nabla}_i^R \right],
\end{equation}
where $\boldsymbol{\nabla}_i^R \equiv \mathbf{q}_i \times \partial / \partial \mathbf{q}_i$.
The evolution of any observable $\mathcal{O}$, defined as the expectation of a microscopic observable $\hat{\mathcal{O}}$, can be found using $\mathcal{L}^{\dagger}$ as $\dot{\mathcal{O}} = \left \langle \mathcal{L}^{\dagger} \hat{\mathcal{O}} \right \rangle$ where $\langle \rangle$ indicates an expectation over $P_N$.

We consider two microscopic observables: the density, $\hat{\rho} (\mathbf{x}) \equiv  \sum_i^N \delta(\mathbf{r}_i - \mathbf{x}) $, and crystallinity, $\hat{\psi}_V (\mathbf{x}) \equiv \sum_i^N \psi_i ( \mathbf{r}^N) \delta(\mathbf{r}_i - \mathbf{x})$ where $\psi_i$ is a measure of the local crystallinity around particle $i$ which generally depends on the difference in position of particle $i$ with those of its neighbors.
Using $\mathcal{L}^{\dagger}$, we find the respective dynamics of the expectation of $\hat{\rho}$ and $\hat{\psi}$ to be $\dot{\rho} = - \boldsymbol{\nabla} \cdot ( U_0 \mathbf{m} + \boldsymbol{\sigma}_C/\zeta)$ and $\dot{\psi}_V = U_0 m^{\psi} + s_C$ after discarding the flux of $\psi_V$ (consistent with Ref.~\cite{Evans2024}).
We have introduced the conservative stress (with Korteweg-like interfacial contributions~\cite{Korteweg1904}) and crystallinity generation, $\boldsymbol{\sigma}_C$ and $s_C$ respectively, which each arise from conservative interparticle interactions. 
We have also introduced the polarization field $\mathbf{m}(\mathbf{x}) \equiv \left \langle \sum_i^N \mathbf{q}_i \delta(\mathbf{r}_i - \mathbf{x}) \right \rangle$ and analogous $m^{\psi}(\mathbf{x}) \equiv \left \langle \sum_i^N \mathbf{q}_j \cdot \partial \psi_i / \partial \mathbf{r}_j \delta(\mathbf{r}_i - \mathbf{x}) \right \rangle$.

Importantly, $\mathbf{m}$ and $m^{\psi}$ each obey their own evolution equation (found by applying $\mathcal{L}^{\dagger}$ to their microscopic definitions) which introduces the nematic order field in $\dot{\mathbf{m}}$ and analogous terms in $\dot{m}^{\psi}$, where these fields again obey their own evolution equation.
Generally, the dynamics of a field depend on at least one other field whose microscopic definition contains an additional factor of orientation.
This infinite hierarchy of orientational moments requires the introduction of closures -- here, consistent with our gradient expansion, we close the hierarchy by assuming fields related to third orientational moments are isotropic. 
Furthermore, we assume conservative interactions appearing in $\dot{\mathbf{m}}$ and $\dot{m}^{\psi}$ and higher order orientational moments (i.e.,~not affecting $\dot{\rho}$ and $\dot{\psi}_V$) can be approximated by introducing an equation of state $\overline{U}(\rho, \psi_V)$ (with value between $0$ and $1$) that captures the effective reduction in self-propulsion speed due to interpactice interactions.
With these closures and approximations, we find the steady-state conditions to be:
\begin{subequations}
    \begin{align}
        & \dot{\rho} = 0 = \frac{1}{\zeta} \partial^2_{zz} \mathcal{P} = \frac{1}{\zeta} \partial^2_{zz} \left[ p_C + p^{\rm bulk}_{\rm act} - \frac{\ell^2_0 \overline{U}}{20} \partial_z \left( \overline{U} \partial_z p_C^{\rm bulk}\right) \right], \\
        & \dot{\psi}_V = 0 = M f_{\psi} = s_C - \frac{\ell^2_0}{24} \partial_z  \left[ \overline{U} \partial_z \left(\overline{U} s_C^{\rm bulk} \right) \right] = 0,
    \end{align}
\end{subequations}
and identify $L=1/\zeta$, $f_{\rho}=-\partial_z \mathcal{P}$, and take $M=1$.
We then find $\mathcal{T}_{\rho \rho} = -1$ and $\mathcal{T}_{\rho \psi} = 0$ and identify $u_{\rho}$ and $u_{\psi}$ [see Eq.~\eqref{eq:pseudopotentials}].

\begin{figure}
	\centering
	\includegraphics[width=.475\textwidth]{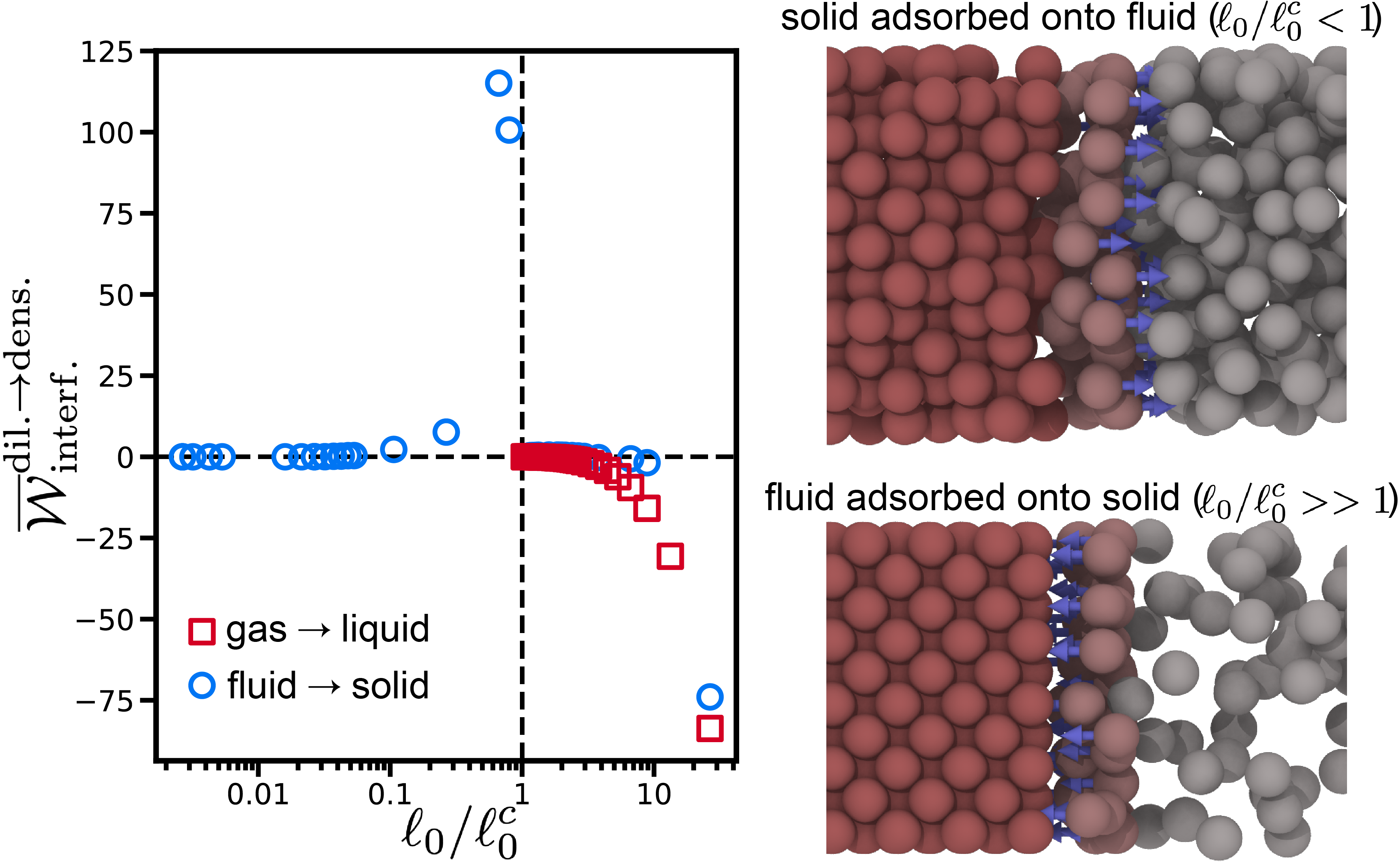}
	\caption{\protect\small{{Dimensionless work (nondimensionalized by the hard sphere energy scale ${\zeta U_0 D}$) to move a particle across the interface during coexistence. Schematics depict the transition of the active force density within the interface from pointing into the fluid phase at low activity (top) to pointing into the solid phase at high activity (bottom).}}}
	\label{fig:work}
\end{figure}

\fakesection{Interfacial Work and Polarization} \label{secSI:work}
\textit{Appendix II: Interfacial Work and Polarization.--} The degree to which the equilibrium Gibbs-Duhem relation is violated can provide direct insight into the nature of the interface dividing two coexisting phases. 
We \textit{define} the work required to move a particle across the interface from the dilute phase (gas/fluid) into the dense phase (liquid/solid) as~\cite{Omar2023b}:
\begin{equation}
    \label{eq:Wdef}
    \mathcal{W}^{\rm dil. \rightarrow dens.}_{\rm interf} \equiv \int_{\upsilon^{\rm dil.}}^{\upsilon^{\rm dens.}} \left[ \mathcal{P}^{\rm bulk} \left( \phi, \psi_V^* \right) - \mathcal{P}^{\rm coexist} \right] d \upsilon,
\end{equation}
where this work is identically zero when the equilibrium Gibbs-Duhem relation is recovered.
We compute this insertion work for both liquid-gas and solid-fluid coexistence, as shown in Fig.~\ref{fig:work}.
For all activities, work is required to move a particle from the liquid phase into the gas phase (${\mathcal{W}^{\rm gas \rightarrow liq.}_{\rm interf} \leq 0}$), as reported in Ref.~\cite{Omar2023b}.
It is only at the critical point, where the ``phases'' are indistinguishable, that the work is identically zero.

\begin{figure}
	\centering
	\includegraphics[width=.35\textwidth]{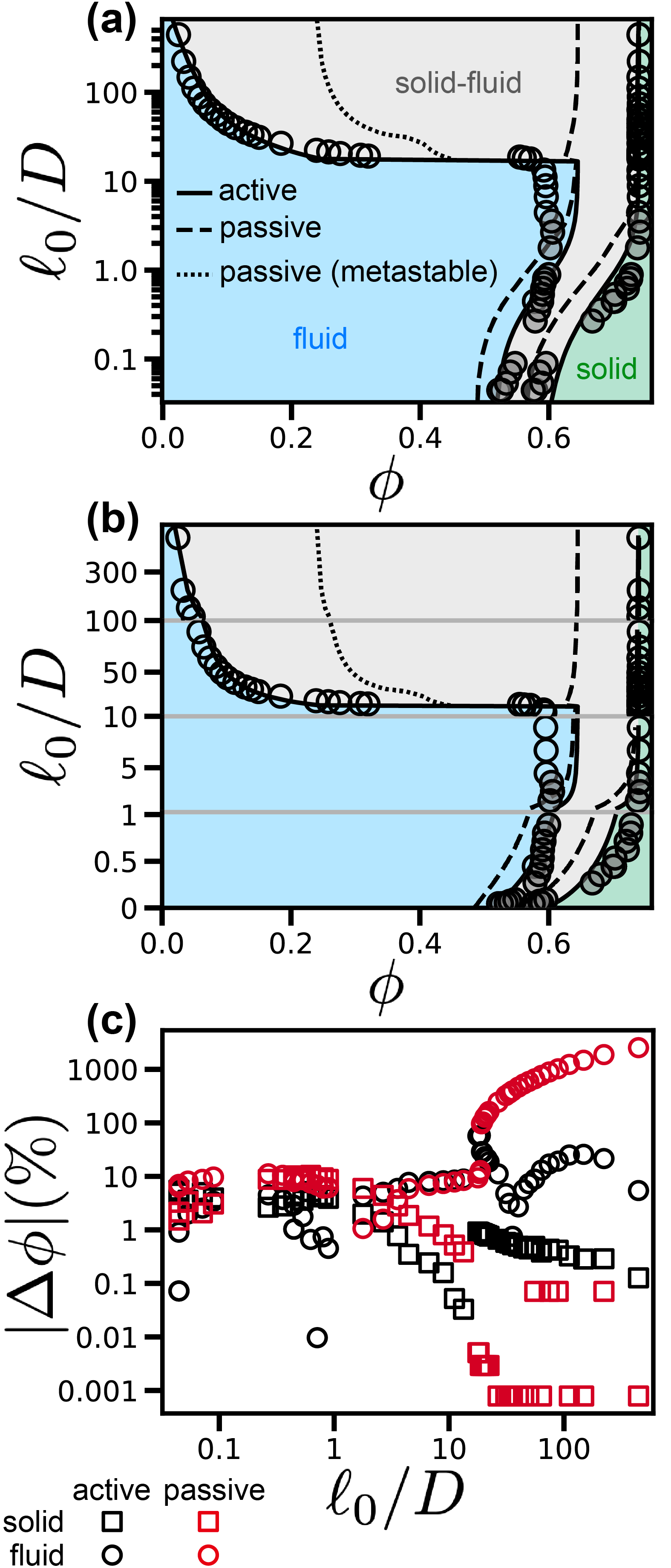}
	\caption{\protect\small{{Solid-fluid phase diagram of 3D active hard spheres on (a) a logarithmic scale and (b) a series of linear scales. The result using the equilibrium (passive) criteria is shown in dashed lines, with the metastable branch shown in dotted lines while the result using the nonequilibrium (active) criteria is shown in solid lines. See Ref.~\cite{Omar2023b} for an analogous comparison of the predictions for the liquid-gas binodal. (c) The percent error of the predicted solid-gas phase diagram using both the equilibrium and nonequilibrium criteria.}}}
	\label{sfig:pdSI}
\end{figure}

The physical origin of this required non-zero insertion work is the polarization of active particles within the interface: active particles within the liquid-gas interface are oriented towards the liquid phase, generating an active force density (see schematic in Fig.~\ref{fig:work}). 
The presence of this force density is \textit{required} for the two phases to mechanically coexist with one another.
The direction of this force density is towards the phase with the lower active pressure which, in the case of disordered active hard sphere fluids, is \textit{always} the denser phase (i.e., the liquid). 
This interfacial force density -- which is not present for passive systems -- must be overcome when a particle is moved out of the liquid phase. 

In the case of solid-fluid coexistence, the insertion work vanishes in the reversible limit (${{\ell_0/D\rightarrow 0}}$) (see Fig.~\ref{fig:work}), consistent with the recovery of the equilibrium crystallization transition.
Departing from the equilibrium limit, we observe that the work required to move a particle from the solid phase into the fluid phase is \textit{negative} despite the solid having the higher density of the two phases.
At low activities (below the triple point), the density contrast between solid and fluid is relatively small (see Fig.~\ref{fig:pd}).
Despite the slightly higher density, the crystalline solid results in more free volume available to the particles in comparison to the dense disordered fluid, resulting in the solid exhibiting a \textit{higher} active pressure than the fluid. 
This causes the force density to point towards the less dense fluid and makes the insertion work negative, shown schematically in Fig.~\ref{fig:work}.
Above the triple point activity, the fluid density markedly decreases, ultimately reversing the sign of the insertion work at high activities. 

\fakesection{Active Phase Diagram Using Equilibrium Coexistence Criteria} \label{secSI:comparecriteria}

\textit{Appendix III: Active Phase Diagram Using Equilibrium Coexistence Criteria.--} While the equilibrium Maxwell construction on $\mathcal{P}^{\rm bulk}$ is both path-dependent and generally finite at nonzero activity, we may still evaluate it along $\psi_V^*$ (Eq.~\eqref{eq:active_criteria_approx} using $\mathcal{E}_{\rho}^{\rm int} = \rho$) to obtain a thermodynamically constructed ABP phase diagram.
We do this in Fig.~\ref{sfig:pdSI}, comparing the solid-fluid phase diagram predicted using our nonequilibrium criteria (Eq.~\eqref{eq:active_criteria_approx} using $\mathcal{E}_{\rho}^{\rm int} = p_C^{\rm bulk}$) with that predicted using the equilibrium criteria.
It is clear from Fig.~\ref{sfig:pdSI}(c) that the two theories provide similar accuracy at low and intermediate activities and both capture the close-packed solid density at high-activity.
However, the fluid densities predicted by the nonequilibrium criteria are significantly more accurate at high activity, where the equilibrium theory overpredicts the fluid density by $\sim 1000 \%$ while the nonequilibrium theory overpredicts the fluid density by only $\sim 1$-$10\%$.

The discrepancy between the active and passive predictions is partially due to the fact that the equilibrium criteria, which amount to minimizing the free energy density ${f^{\rm bulk}(\rho, \psi_V^*) \equiv -\mathcal{P}^{\rm bulk}(\rho, \psi_V^*) + \rho \int \upsilon d \mathcal{P}^{\rm bulk}(\rho, \psi_V^*)}$ subject to fixed total density and system volume, predict that the solid phase coexists with a liquid phase whose density exceeds the MIPS binodal.
Solid-gas coexistence has higher free energy than MIPS and is thus metastable, leaving solid-liquid coexistence as the globally stable form of solid-fluid coexistence predicted by the passive criteria.
This demonstrates that while the equilibrium construction can still be used at low activities (as this is precisely the reversible limit), its erroneous use can cause significant qualitative inaccuracies at finite activities.

\end{document}


\title{Supplemental Material -- Theory of Nonequilibrium Crystallization and the Phase Diagram of Active Brownian Spheres}
\author{Daniel Evans}
\author{Ahmad K. Omar}
\email{aomar@berkeley.edu}
\affiliation{Department of Materials Science and Engineering, University of California, Berkeley, California 94720, USA}
\affiliation{Materials Sciences Division, Lawrence Berkeley National Laboratory, Berkeley, California 94720, USA}

\maketitle
\tableofcontents

\section{Microscopic Derivation of the Evolution of the Density and Crystallinity Fields}
In this Section, we derive the dynamics of the ensemble-averaged density and crystallinity fields to identify the steady-state force balance governing each field.
We derive the density field dynamics in an identical manner as in Ref.~\cite{Omar2023b} and use a similar approach to derive the crystallinity field dynamics.
\subsection{Equations of Motion and Fokker-Planck Operator}
We consider $N$ interacting 3D active Brownian particles, where the position, $\mathbf{r}_i$, and orientation, $\mathbf{q}_i$, of the $i^{th}$ particle undergo the following equations of motion:
\begin{subequations}
    \begin{align}
        & \dot{\mathbf{r}}_i = U_0 \mathbf{q}_i + \frac{1}{\zeta} \mathbf{F}_i^C, \\
        & \dot{\mathbf{q}}_i = \boldsymbol{\Omega}_i \times \mathbf{q}_i,
    \end{align}
\end{subequations}
where $\dot{a} \equiv \partial a / \partial t$, $U_0$ is the active speed, $\zeta$ is the drag coefficient, $\mathbf{F}_i^C=\sum_{j \neq i} \mathbf{F}_{ij}$ is the sum of conservative interparticle forces on particle $i$, and $\boldsymbol{\Omega}_i$ is a random angular velocity with zero mean and variance $\left \langle \boldsymbol{\Omega}_i(t) \boldsymbol{\Omega}_j(t') \right \rangle = 2 \tau_R^{-1} \delta_{ij} \delta(t-t') \mathbf{I}$ where $\tau_R$ is the characteristic reorientation time, $\delta_{ij}$ is the Kronecker delta, $\delta(t)$ is the Dirac delta function, and $\mathbf{I}$ is the identity tensor.
Here, this mean and variance are taken over the noise statistics.

The noise-averaged $N$-body distribution function, $P_N$, evolves according to $\dot{P}_N = \mathcal{L} P_N$ where the Fokker-Planck operator, $\mathcal{L}$, is:
\begin{equation}
    \mathcal{L} \equiv -\sum_i \left[ \frac{\partial}{\partial \mathbf{r}_i} \cdot \left( U_0 \mathbf{q}_i + \frac{1}{\zeta} \mathbf{F}_i^C \right) - \tau_R^{-1} \boldsymbol{\nabla}_i^R \cdot \boldsymbol{\nabla}_i^R \right],
\end{equation}
where $\boldsymbol{\nabla}_i^R \equiv \mathbf{q}_i \times \partial / \partial \mathbf{q}_i$ is the rotational gradient operator.
The adjoint to this Fokker-Planck operator, $\mathcal{L}^{\dagger}$, is:
\begin{equation}
    \mathcal{L}^{\dagger} \equiv \sum_i \left[  \left( U_0 \mathbf{q}_i + \frac{1}{\zeta} \mathbf{F}_i^C \right) \cdot \frac{\partial}{\partial \mathbf{r}_i} + \tau_R^{-1} \boldsymbol{\nabla}_i^R \cdot \boldsymbol{\nabla}_i^R \right].
\end{equation}
This adjoint allows one to express the evolution of an observable $\mathcal{O} \equiv \left \langle \hat{\mathcal{O}} \right \rangle$ as $\dot{\mathcal{O}} = \left \langle \mathcal{L}^{\dagger} \hat{\mathcal{O}} \right \rangle$, where $\hat{\mathcal{O}}$ is the microscopic definition of the observable.
Here, expectations are now taken over the $N$-body distribution, $\left \langle \hat{\mathcal{O}} \right \rangle = \int_{\gamma} d \boldsymbol{\Gamma} P_N \hat{\mathcal{O}}$ where $\boldsymbol{\Gamma} = \{ \mathbf{r}^N, \mathbf{q}^N \}$ contains all $N$ positions and $N$ orientations and $\gamma$ is the phase space volume.

\subsection{Steady-State Mechanics of the Density Field}
We now consider the evolution of the density field, $\rho(\mathbf{x}, t) \equiv \left \langle \sum_i \delta(\mathbf{x} - \mathbf{r}_i) \right \rangle$.
This derivation of the density field dynamics was previously performed in Ref.~\cite{Omar2023b}, using the same closures, approximations, and constitutive relations that we employ here.
While a detailed discussion of this derivation and the nature of the approximations and closures can be found in Ref.~\cite{Omar2023b}, we briefly recapitulate the essence of the derivation here for convenience.

Using $\mathcal{L}^{\dagger}$, we find the dynamics of $\rho$ to be:
\begin{equation}
    \label{eq:rhoevolution}
    \dot{\rho} = \left \langle \mathcal{L}^{\dagger} \sum_i \delta(\mathbf{x} - \mathbf{r}_i) \right \rangle = - \boldsymbol{\nabla} \cdot \left( U_0 \mathbf{m} + \frac{1}{\zeta} \boldsymbol{\nabla} \cdot \boldsymbol{\sigma}_C \right),
\end{equation}
where we have defined the polar order field, $\mathbf{m}(\mathbf{x}, t) \equiv \left \langle \sum_i \mathbf{q}_i \delta(\mathbf{x} - \mathbf{r}_i) \right \rangle$, and the conservative stress, $\boldsymbol{\sigma}_C \equiv \frac{1}{2} \left \langle \sum_i \sum_{j \neq i} \mathbf{r}_{ij} \mathbf{F}_{ij} b_{ij} \right \rangle$ with the distance vector $\mathbf{r}_{ij} \equiv \mathbf{r}_j - \mathbf{r}_i$ and bond function $b_{ij} \equiv \int_0^1 d \lambda \delta(\mathbf{x} - \mathbf{r}_j + \lambda \mathbf{r}_{ij})$.

\sloppy The polar order field obeys its own evolution equation.
Before examining the resulting equation, we introduce an approximation in the dynamics of \textit{all} fields whose microscopic definition includes at least one orientation: the conservative interactions act to reduce the effective active speed from its ideal value $U_0$ to a field-dependent value. 
This dependence will ultimately require additional constitutive equations (see Ref.~\cite{Omar2023b}).
Using this approximation, the evolution of the polar order field is:
\begin{equation}
    \label{eq:mevolution}
    \dot{\mathbf{m}} = \left \langle \mathcal{L}^{\dagger} \sum_i \mathbf{q}_i \delta(\mathbf{x} - \mathbf{r}_i) \right \rangle = - \boldsymbol{\nabla} \cdot \left( U_0 \overline{U}^{\mathbf{m}} \tilde{\mathbf{Q}} \right) - \frac{2}{\tau_R} \mathbf{m},
\end{equation}
where we have introduced the nematic order field $\tilde{\mathbf{Q}} \equiv \left \langle \sum_i \mathbf{q}_i \mathbf{q}_i \delta( \mathbf{x} - \mathbf{r}_i) \right \rangle$ and the renormalized active speed of the polar order field $U_0 \overline{U}^{\mathbf{m}}$.
The nematic order field undergoes its own evolution equation:
\begin{equation}
    \dot{\tilde{\mathbf{Q}}} = \left \langle \mathcal{L}^{\dagger} \sum_i \mathbf{q}_i \mathbf{q}_i \delta(\mathbf{x} - \mathbf{r}_i) \right \rangle = - \boldsymbol{\nabla} \cdot \left( U_0 \overline{U}^{\tilde{\mathbf{Q}}} \tilde{\mathbf{B}} \right) - \frac{6}{\tau_R} \left( \tilde{\mathbf{Q}} - \frac{\rho}{3} \mathbf{I} \right),
\end{equation}
where we have introduced the third orientational moment field, ${\tilde{\mathbf{B}} \equiv \left \langle \sum_i \mathbf{q}_i \mathbf{q}_i \mathbf{q}_i \delta( \mathbf{x} - \mathbf{r}_i) \right \rangle}$, and the renormalized active speed of the nematic field $U_0 \overline{U}^{\mathbf{Q}}$.
Notably, the traceless nematic order parameter, defined as $ \mathbf{Q} \equiv \tilde{\mathbf{Q}} - \rho \mathbf{I} / 3$, has the following dynamics:
\begin{equation}
    \dot{\mathbf{Q}}= - \boldsymbol{\nabla} \cdot \left( U_0 \overline{U}^{\tilde{\mathbf{Q}}} \tilde{\mathbf{B}} \right) - \frac{1}{3} \mathbf{I} \dot{\rho} - \frac{6}{\tau_R} \mathbf{Q} = - \boldsymbol{\nabla} \cdot \left( U_0 \overline{U}^{\tilde{\mathbf{Q}}} \tilde{\mathbf{B}} - \frac{1}{3} \mathbf{I} U_0 \mathbf{m} - \frac{1}{3 \zeta} \boldsymbol{\nabla} \cdot \boldsymbol{\sigma}_C \right) - \frac{6}{\tau_R} \mathbf{Q}.
\end{equation}

While $\tilde{\mathbf{B}}$ undergoes its own equation of motion, we define $\tilde{\mathbf{B}} = \mathbf{B} + \boldsymbol{\alpha} \cdot \mathbf{m} / 5$ (where $\boldsymbol{\alpha}$ is an isotropic fourth rank tensor with $\alpha_{ijkl} = \delta_{ij} \delta_{kl} + \delta_{ik} \delta_{jl} + \delta_{il} \delta_{jk}$ in Einstein notation) and use the closure $\mathbf{B}=\mathbf{0}$. 
This closure allows us to simply express the third orientational moment with the polar order field.
The evolution of the nematic order field is then:
\begin{equation}
    \label{eq:Qevolution}
    \dot{\mathbf{Q}} = - \boldsymbol{\nabla} \cdot \left( U_0 \mathbf{I} \mathbf{m} \left( \frac{3}{5}\overline{U}^{\tilde{\mathbf{Q}}}  - \frac{1}{3} \right) - \frac{1}{3 \zeta} \boldsymbol{\nabla} \cdot \boldsymbol{\sigma}_C \right) - \frac{6}{\tau_R} \mathbf{Q}.
\end{equation}

Inserting the steady-state solution to Eq.~\eqref{eq:mevolution} into Eq.~\eqref{eq:rhoevolution} we have:
\begin{equation}
    \dot{\rho} = - \boldsymbol{\nabla} \cdot \left( \frac{1}{\zeta} \boldsymbol{\nabla} \cdot \boldsymbol{\sigma}_{\rm act} + \frac{1}{\zeta} \boldsymbol{\nabla} \cdot \boldsymbol{\sigma}_C \right),
\end{equation}
where we have defined the active stress $ \boldsymbol{\sigma}_{\rm act} \equiv -\zeta U_0 \ell_0 \overline{U}^{\mathbf{m}} \tilde{\mathbf{Q}} / 2$ with the run length $\ell_0 \equiv U_0 \tau_R$.

We now explicitly look for the quasi-1D dynamics in the $z$-direction.
Defining $\mathcal{P} \equiv -\sigma^{zz}_{\rm act} + p_C$ where $p_C \equiv - \sigma^{zz}_C$ and assuming the polar and nematic order fields relax faster than the density field (quasi-static density field dynamics), 
we find:
\begin{subequations}
\begin{equation}
    \dot{\rho} = \frac{\partial }{\partial z} \left( \frac{1}{\zeta} \frac{\partial }{\partial z} \mathcal{P}\right) = - \frac{\partial}{\partial z} \left( L f_{\rho} \right),
\end{equation}
and identify $J_{\rho} = L f_{\rho} = - \zeta \partial \mathcal{P} / \partial z$, mapping these dynamics to Ref.~\cite{Evans2024}.
We see that $L = \zeta^{-1}$ and therefore find the flux-driving force to be:
\begin{equation}
    \label{eq:frho}
    f_{\rho} = -\frac{\partial }{\partial z} \mathcal{P} = -\frac{\partial }{\partial z} \left[ p_C + p^{\rm bulk}_{\rm act} - \frac{\ell^2_0 \overline{U}}{20} \frac{\partial }{\partial z} \left( \overline{U} \frac{\partial  p_C^{\rm bulk}}{\partial  z}\right) \right],
\end{equation}
\end{subequations}
where we have defined the active pressure $p^{\rm bulk}_{\rm act} \equiv \zeta U_0 \ell_0 \overline{U} \rho / 6$ and, in line with our quasi-static approximation, have substituted the steady-state solution to Eq.~\eqref{eq:Qevolution}, the steady-state solution $\zeta U_0 m_z = - d \sigma^{zz} / dz$, and the approximation $\overline{U}^{\mathbf{m}} = \overline{U}^{\tilde{\mathbf{Q}}} = \overline{U}$ into $\mathcal{P}$ and truncated at second order in spatial gradients.
Equation~\eqref{eq:frho} immediately yields the ``chemical pseudopotential'' $u_{\rho}=\mathcal{P}$ with $\mathcal{T}_{\rho \rho} = -1$ and $\mathcal{T}_{\rho \psi} = 0$, noting that $\partial / \partial z \rightarrow d / dz$ in a steady-state.

\subsection{Steady-State Mechanics of the Crystallinity Field}
We now consider the evolution of a crystallinity field, $\psi_V(\mathbf{x}, t) \equiv \left \langle \sum_i \psi_i \delta(\mathbf{x} - \mathbf{r}_i) \right \rangle$, where $\psi_i$ provides a measure of the local crystalline order around particle $i$ that in principle can depend on the distance vectors $\mathbf{r}_{ij}$ between $i$ and all other particles $j$.
This derivation has not been performed previously to our knowledge, however we aim to parallel the derivation of the density field dynamics when possible.
Using $\mathcal{L}^{\dagger}$ we find:
\begin{equation}
    \label{eq:psievolution}
    \dot{\psi}_V = \left \langle \mathcal{L}^{\dagger} \sum_i \psi_i \delta(\mathbf{x} - \mathbf{r}_i) \right \rangle = - \boldsymbol{\nabla} \cdot \mathbf{J}_{\psi} +  U_0 m^{\psi} + s_C,
\end{equation}
where we have defined $m^{\psi} \equiv \left \langle \sum_i \sum_j \mathbf{q}_j \cdot \frac{\partial \psi_i}{\partial \mathbf{r_j}} \delta (\mathbf{x} - \mathbf{r}_i) \right \rangle$, $s_C \equiv \frac{1}{\zeta} \left \langle \sum_i \sum_j \mathbf{F}_j^C \cdot \frac{\partial \psi_i}{\partial \mathbf{r_j}} \delta (\mathbf{x} - \mathbf{r}_i) \right \rangle$, and the crystallinity flux $\mathbf{J}_{\psi}$:
\begin{equation}
    \mathbf{J}_{\psi} \equiv \left \langle \sum_i \left( U_0 \mathbf{q}_i + \frac{1}{\zeta} \mathbf{F}^C_i\right) \delta \left( \mathbf{x} - \mathbf{r}_i \right) \right \rangle.
\end{equation}
While the flux is clearly not identically zero, we nevertheless approximate it as negligible in comparison to the generation terms, $|\boldsymbol{\nabla} \cdot \mathbf{J}_{\psi}| << |s_{\psi}|$. 
This results in model A dynamics for the crystallinity field and allows us to apply our recently proposed coexistence framework~\cite{Evans2024}.

As was the case in the previous section, $m^{\psi}$ undergoes its own evolution equation:
\begin{equation}
    \label{eq:mpsievolution}
    \dot{m}^{\psi} = \left \langle \mathcal{L}^{\dagger} \sum_i \sum_j \mathbf{q}_j \cdot \frac{\partial \psi_i}{\partial \mathbf{r_j}} \delta (\mathbf{x} - \mathbf{r}_i) \right \rangle = - \boldsymbol{\nabla} \cdot \left( U_0 \overline{U}^{m^{\psi}}_{\mathbf{J}} \tilde{\mathbf{Q}}^{\psi}_1 \right) + U_0 \overline{U}^{m^{\psi}}_{s} \tilde{Q}^{\psi}_2 - \frac{2}{\tau_R} m^{\psi},
\end{equation}
where we have defined:
\begin{subequations}
    \begin{align}
        & \tilde{\mathbf{Q}}^{\psi}_1 = \left \langle \sum_i \sum_j \mathbf{q}_i \mathbf{q}_j \cdot \frac{\partial \psi_i}{\partial \mathbf{r}_j} \delta(\mathbf{x} - \mathbf{r}_i) \right \rangle, \\
        & \tilde{Q}^{\psi}_2 = \left \langle \sum_i \sum_j \sum_k \mathbf{q}_j \cdot \frac{\partial^2 \psi_i}{\partial \mathbf{r}_j \partial \mathbf{r}_k} \cdot \mathbf{q}_k \delta(\mathbf{x} - \mathbf{r}_i) \right \rangle,
    \end{align}
\end{subequations}
and the renormalized active speeds $U_0 \overline{U}^{m^{\psi}}_{\mathbf{J}}$ and $U_0 \overline{U}^{m^{\psi}}_{s}$.

We now seek expressions for the evolution of $\tilde{\mathbf{Q}}^{\psi}_1$ and $\tilde{Q}^{\psi}_2$.
Beginning with $\tilde{Q}^{\psi}_2$:
\begin{multline}
    \dot{\tilde{Q}}^{\psi}_2 = \left \langle \mathcal{L}^{\dagger} \sum_i \sum_j \sum_k \mathbf{q}_j \cdot \frac{\partial^2 \psi_i}{\partial \mathbf{r}_j \partial \mathbf{r}_k} \cdot \mathbf{q}_k \delta(\mathbf{x} - \mathbf{r}_i) \right \rangle \\ = - \boldsymbol{\nabla} \cdot \left( U_0 \overline{U}_{\mathbf{J}}^{\tilde{Q}^{\psi}_2} \tilde{\mathbf{B}}_2 \right) + U_0 \overline{U}_{s}^{\tilde{Q}^{\psi}_2} \tilde{B}_3 - \frac{4}{\tau_R} \tilde{Q}^{\psi}_2 + \frac{2}{\tau_R} \left \langle \sum_i \sum_j \sum_k \frac{\partial}{\partial \mathbf{r}_j} \cdot \frac{\partial \psi_i}{\partial \mathbf{r}_k}\delta(\mathbf{x} - \mathbf{r}_i) \right \rangle \\ - \frac{1}{\tau_R} \left \langle \sum_i \sum_j \sum_k \left( \mathbf{q}_j \mathbf{q}_j + \mathbf{q}_k \mathbf{q}_k \right) : \frac{\partial^2 \psi_i}{\partial \mathbf{r}_j \partial \mathbf{r}_k} \delta(\mathbf{x} - \mathbf{r}_i) \right \rangle,
\end{multline}
where we have defined:
\begin{subequations}
    \begin{align}
        & \tilde{\mathbf{B}}^{\psi}_2 = \left \langle \sum_i \sum_j \sum_k \mathbf{q}_i \mathbf{q}_j \cdot \frac{\partial^2 \psi_i}{\partial \mathbf{r}_j \partial \mathbf{r}_k} \cdot \mathbf{q}_k \delta(\mathbf{x} - \mathbf{r}_i) \right \rangle, \\
        & \tilde{B}^{\psi}_3 = \left \langle \sum_i \sum_j \sum_k \sum_l \mathbf{q}_j \mathbf{q}_k \mathbf{q}_l \vdots \frac{\partial^3 \psi_i}{\partial \mathbf{r}_j \partial \mathbf{r}_k \partial \mathbf{r}_l} \delta(\mathbf{x} - \mathbf{r}_i) \right \rangle,
    \end{align}
\end{subequations}
along with the renormalized active speeds $U_0 \overline{U}_{\mathbf{J}}^{\tilde{Q}^{\psi}_2}$ and $ U_0 \overline{U}_{s}^{\tilde{Q}^{\psi}_2}$.
We now recognize that as each $\psi_i$ is a function of the bond vectors $\mathbf{r}_{ij}$, $\partial \psi_i / \partial \mathbf{r}_i = - \sum_{j \neq i} \partial \psi_i / \partial \mathbf{r}_{ij}$ and $\partial \psi_i / \partial \mathbf{r}_j = \partial \psi_i / \partial \mathbf{r}_{ij}$.
This implies:
\begin{subequations}
    \begin{equation}
        \left \langle \sum_i \sum_j \sum_k \frac{\partial}{\partial \mathbf{r}_j} \cdot \frac{\partial \psi_i}{\partial \mathbf{r}_k}\delta(\mathbf{x} - \mathbf{r}_i) \right \rangle = \left \langle \sum_i \sum_j  \frac{\partial}{\partial \mathbf{r}_j} \cdot \left(  \frac{\partial \psi_i}{\partial \mathbf{r}_i} + \sum_{k \neq i} \frac{\partial \psi_i}{\partial \mathbf{r}_k} \right) \delta(\mathbf{x} - \mathbf{r}_i) \right \rangle = 0,
    \end{equation}
    and:
    \begin{multline}
        \left \langle \sum_i \sum_j \sum_k \left( \mathbf{q}_j \mathbf{q}_j + \mathbf{q}_k \mathbf{q}_k \right) : \frac{\partial^2 \psi_i}{\partial \mathbf{r}_j \partial \mathbf{r}_k} \delta(\mathbf{x} - \mathbf{r}_i) \right \rangle = \\ \bigg \langle \sum_i \bigg( \sum_j \mathbf{q}_j \mathbf{q}_j  : \frac{\partial}{\partial \mathbf{r}_j} \left[ \frac{\partial \psi_i}{ \partial \mathbf{r}_i} + \sum_{k \neq i} \frac{\partial \psi_i}{ \partial \mathbf{r}_k} \right] \\ + \sum_k \mathbf{q}_k \mathbf{q}_k  : \frac{\partial}{\partial \mathbf{r}_k} \left[ \frac{\partial \psi_i}{ \partial \mathbf{r}_i} + \sum_{j \neq i} \frac{\partial \psi_j}{ \partial \mathbf{r}_k} \right] \bigg) \delta(\mathbf{x} - \mathbf{r}_i) \bigg \rangle = 0,
    \end{multline}
\end{subequations}
and hence:
\begin{equation}
   \label{eq:Q2evolutionpre}
    \dot{\tilde{Q}}^{\psi}_2 = - \boldsymbol{\nabla} \cdot \left( U_0 \overline{U}_{\mathbf{J}}^{\tilde{Q}^{\psi}_2} \tilde{\mathbf{B}}_2 \right) + U_0 \overline{U}_{s}^{\tilde{Q}^{\psi}_2} \tilde{B}_3 - \frac{4}{\tau_R} \tilde{Q}^{\psi}_2.
\end{equation}

We examine the evolution of $\tilde{\mathbf{Q}}^{\psi}_1$:
\begin{multline}
    \dot{\tilde{\mathbf{Q}}}^{\psi}_1 = \left \langle \mathcal{L}^{\dagger} \sum_i \sum_j \mathbf{q}_i \mathbf{q}_j \cdot \frac{\partial\psi_i}{\partial \mathbf{r}_j} \delta(\mathbf{x} - \mathbf{r}_i) \right \rangle \\ = - \boldsymbol{\nabla} \cdot \left( U_0 \overline{U}_{\mathbf{J}}^{\tilde{\mathbf{Q}}^{\psi}_1} \tilde{\mathbf{B}}_1 \right) + U_0 \overline{U}_{s}^{\tilde{\mathbf{Q}}^{\psi}_1} \tilde{\mathbf{B}}_2 - \frac{4}{\tau_R} \tilde{\mathbf{Q}}^{\psi}_1 + \frac{2}{\tau_R} \left \langle \sum_i \sum_j \frac{\partial \psi_i}{\partial \mathbf{r}_j}\delta(\mathbf{x} - \mathbf{r}_i) \right \rangle \\ - \frac{1}{\tau_R} \left \langle \sum_i \sum_j \left( \mathbf{q}_j \mathbf{q}_j + \mathbf{q}_i \mathbf{q}_i \right) \cdot \frac{\partial\psi_i}{\partial \mathbf{r}_j} \delta(\mathbf{x} - \mathbf{r}_i) \right \rangle,
\end{multline}
where we have defined:
\begin{equation}
    \tilde{\mathbf{B}}^{\psi}_1 = \left \langle \sum_i \sum_j  \mathbf{q}_i \mathbf{q}_i \mathbf{q}_j \cdot \frac{\partial \psi_i}{\partial \mathbf{r}_j} \delta(\mathbf{x} - \mathbf{r}_i) \right \rangle,
\end{equation}
and the renormalized active speeds $U_0 \overline{U}_{\mathbf{J}}^{\tilde{Q}^{\psi}_1}$ and $ U_0 \overline{U}_{s}^{\tilde{Q}^{\psi}_1}$.
We again find:
\begin{equation}
    \left \langle \sum_i \sum_j \frac{\partial \psi_i}{\partial \mathbf{r}_j}\delta(\mathbf{x} - \mathbf{r}_i) \right \rangle = \left \langle \sum_i \left( \frac{\partial \psi_i}{\partial \mathbf{r}_i} +  \sum_{j \neq i} \frac{\partial \psi_i}{\partial \mathbf{r}_j} \right) \delta(\mathbf{x} - \mathbf{r}_i) \right \rangle =0.
\end{equation}
Contrasting this, we find:
\begin{align}
    \bigg \langle \sum_i \sum_j & \left( \mathbf{q}_j \mathbf{q}_j + \mathbf{q}_i \mathbf{q}_i \right) \cdot \frac{\partial\psi_i}{\partial \mathbf{r}_j}  \delta(\mathbf{x} - \mathbf{r}_i)  \bigg \rangle  \nonumber \\ &= \left \langle \sum_i \left[ 2 \mathbf{q}_i \mathbf{q}_i  \cdot \frac{\partial\psi_i}{\partial \mathbf{r}_i} + \sum_{j \neq i} \left( \mathbf{q}_j \mathbf{q}_j + \mathbf{q}_i \mathbf{q}_i \right) \cdot \frac{\partial\psi_i}{\partial \mathbf{r}_j} \right] \delta(\mathbf{x} - \mathbf{r}_i) \right \rangle \nonumber \\ & = \left \langle \sum_i \sum_{j \neq i} \left( \mathbf{q}_j \mathbf{q}_j - \mathbf{q}_i \mathbf{q}_i \right) \cdot \frac{\partial\psi_i}{\partial \mathbf{r}_{ij}} \delta(\mathbf{x} - \mathbf{r}_i) \right \rangle \neq 0,
\end{align}
which makes one unable to obtain closed expressions for the source of $\psi_V$ at this order.
To circumvent this, we assume that this vector can be expressed as the product of an isotropic second order tensor and $\left \langle \sum_i \sum_j \frac{\partial \psi_i}{\partial \mathbf{r}_j}\delta(\mathbf{x} - \mathbf{r}_i) \right \rangle$ which vanishes.
We then have:
\begin{equation}
   \label{eq:Q1evolutionpre}
    \dot{\tilde{\mathbf{Q}}}^{\psi}_1 = - \boldsymbol{\nabla} \cdot \left( U_0 \overline{U}_{\mathbf{J}}^{\tilde{\mathbf{Q}}^{\psi}_1} \tilde{\mathbf{B}}_1 \right) + U_0 \overline{U}_{s}^{\tilde{\mathbf{Q}}^{\psi}_1} \tilde{\mathbf{B}}_2 - \frac{4}{\tau_R} \tilde{\mathbf{Q}}^{\psi}_1.
\end{equation}

To close these equations, we introduce the following forms for $\tilde{\mathbf{B}}^{\psi}_1$:
\begin{subequations}
    \begin{equation}
        \tilde{\mathbf{B}}^{\psi}_1 = \mathbf{B}^{\psi}_i + \frac{1}{3} \mathbf{I}\left \langle \sum_i \sum_j \mathbf{q}_j \cdot \frac{\partial \psi_i}{\partial \mathbf{r}_j} \delta(\mathbf{x} - \mathbf{r}_i) \right \rangle = \mathbf{B}^{\psi}_1 + \frac{1}{3} \mathbf{I} m^{\psi},
    \end{equation}
    for $\tilde{\mathbf{B}}^{\psi}_2$:
    \begin{multline}
        \tilde{\mathbf{B}}^{\psi}_2 = \mathbf{B}^{\psi}_2 + \left \langle \sum_i \sum_j \sum_k \mathbf{q}_j \cdot \frac{\partial^2 \psi_i}{\partial \mathbf{r}_j \partial \mathbf{r}_k} \delta(\mathbf{x} - \mathbf{r}_i) \right \rangle \\ = \mathbf{B}^{\psi}_2 + \left \langle \sum_i \sum_j \mathbf{q}_j \cdot \frac{\partial}{\partial \mathbf{r}_j} \left(\frac{\partial\psi_i}{\partial \mathbf{r}_i} + \sum_{k \neq i} \frac{\partial\psi_i}{\partial \mathbf{r}_k} \right) \delta(\mathbf{x} - \mathbf{r}_i) \right \rangle = \mathbf{B}^{\psi}_2,
    \end{multline}
    and for $\tilde{B}^{\psi}_3$:
    \begin{multline}
        \tilde{B}^{\psi}_3 = B^{\psi}_3 + \left \langle \sum_i \sum_j \sum_k \sum_l \mathbf{q}_j \cdot \frac{\partial}{\partial \mathbf{r}_j} \frac{\partial}{\partial \mathbf{r}_k} \cdot \frac{\partial \psi_i}{\partial \mathbf{r}_l} \delta(\mathbf{x} - \mathbf{r}_i) \right \rangle \\ = B^{\psi}_3 + \left \langle \sum_i \sum_j \sum_k \mathbf{q}_j \cdot \frac{\partial}{\partial \mathbf{r}_j} \frac{\partial}{\partial \mathbf{r}_k} \cdot  \left( \frac{\partial \psi_i}{\partial \mathbf{r}_i} + \sum_{l \neq i} \frac{\partial \psi_i}{\partial \mathbf{r}_l} \right) \delta(\mathbf{x} - \mathbf{r}_i) \right \rangle = B^{\psi}_3.
    \end{multline}
\end{subequations}
We then introduce the closures $\mathbf{B}^{\psi}_1 = \mathbf{0}$, $\mathbf{B}^{\psi}_2 = \mathbf{0}$, and $B^{\psi}_3=0$ and substitute the result into Eqs.~\eqref{eq:Q2evolutionpre} and \eqref{eq:Q1evolutionpre}, finding:
\begin{subequations}
    \label{eq:Qevolutions}
    \begin{align}
        \label{eq:Q2evolution}
        & \dot{\tilde{Q}}^{\psi}_2 = - \frac{4}{\tau_R} \tilde{Q}^{\psi}_2, \\
        \label{eq:Q1evolution}
        & \dot{\tilde{\mathbf{Q}}}^{\psi}_1 = - \boldsymbol{\nabla}\left( U_0 \overline{U}_{\mathbf{J}}^{\tilde{\mathbf{Q}}^{\psi}_1} \frac{1}{3} m^{\psi} \right)  - \frac{4}{\tau_R} \tilde{\mathbf{Q}}^{\psi}_1.
    \end{align}
\end{subequations}

We now return to Eq.~\eqref{eq:psievolution}.
Solving Eq.~\eqref{eq:mpsievolution} and substituting the result into Eq.~\eqref{eq:psievolution} we have:
\begin{equation}
    \dot{\psi}_V = - \boldsymbol{\nabla} \cdot \left( \frac{U_0 \ell_0 \overline{U}_{\mathbf{J}}^{m^{\psi}}}{2} \tilde{\mathbf{Q}}_1^{\psi} \right) +  \frac{U_0 \ell_0 \overline{U}_{s}^{m^{\psi}}}{2}\tilde{Q}_2^{\psi} + s_C
\end{equation}
Substituting the solutions to Eq.~\eqref{eq:Qevolutions} we find:
\begin{equation}
    \dot{\psi}_V = \boldsymbol{\nabla} \cdot \left( \frac{\ell^2_0 \overline{U}_{\mathbf{J}}^{m^{\psi}}}{24} \boldsymbol{\nabla}\left( \overline{U}_{\mathbf{J}}^{\tilde{\mathbf{Q}}^{\psi}_1} U_0 m^{\psi} \right) \right) + s_C.
\end{equation}
We now use the steady-state solution $U_0 m^{\psi} = - s_C$ to find:
\begin{equation}
    \dot{\psi}_V = s_C - \boldsymbol{\nabla} \cdot \left( \frac{\ell^2_0 \overline{U}_{\mathbf{J}}^{m^{\psi}}}{24} \boldsymbol{\nabla}\left( \overline{U}_{\mathbf{J}}^{\tilde{\mathbf{Q}}^{\psi}_1}  s_C \right) \right).
\end{equation}

Moving to a quasi-1D profile oriented along the $z$-direction, the dynamics of the crystallinity field are:
\begin{equation}
    \dot{\psi}_V = M f_{\psi} = s_C - \frac{\ell^2_0}{24} \frac{\partial }{\partial z} \left( \overline{U} \frac{\partial }{\partial z}\left(\overline{U} s_C^{\rm bulk} \right) \right),
\end{equation}
where we have truncated at second order in spatial gradients by retaining only the bulk part of the conservative generation ($s_C \equiv s_C^{\rm bulk} + s_C^{(2)}$) and assumed $\overline{U}_{\mathbf{J}}^{m^{\psi}} = \overline{U}_{\mathbf{J}}^{\tilde{\mathbf{Q}}^{\psi}_1} = \overline{U}$.
Mapping these dynamics to Ref.~\cite{Evans2024}, we may set $M=1$ and identify the generation-driving force as:
\begin{equation}
    f_{\psi} = s_C - \frac{\ell^2_0}{24} \frac{\partial }{\partial z} \left( \overline{U} \frac{\partial }{\partial z}\left(\overline{U} s_C^{\rm bulk} \right) \right),
\end{equation}
where $\partial / \partial z \rightarrow d / d z$ in a steady-state.

\section{Equations of State of Active Brownian Spheres}
\label{secSI:eos}
Ultimately, the application of the coexistence criteria derived in the main text will require equations of state for $p_C^{\rm bulk}$, $\overline{U}$, and $\psi^*_V$.
Simulation data can \textit{only} be obtained for systems in which a state of homogeneous $\rho$ is at least locally stable.
Consequently, it is not possible to obtain the complete relevant functional dependence of the required state functions directly from simulation. 
However, application of our coexistence criteria only requires knowledge of the equations of state at $\psi_V = \psi_V^*(\rho)$ for each density $\rho$.
We therefore proceed by devising a simple simulation protocol to obtain as much of this limited data as possible. 
We then use this data, along with the known physical limits we require our equations of state to capture, in order to develop physical and semi-empirical bulk equations of state.

\subsection{\label{secSI:simdetails}Simulation Details}

Brownian dynamics simulations (see main text) of active hard spheres were performed following Ref.~\cite{Omar2021pd}.
The hard-sphere diameter, $D$, is the only natural length scale in addition to the run length. 
As a result, the system state is entirely characterized by two dimensionless, intensive, geometric parameters: the volume fraction of spheres $\phi \equiv \rho \pi D^3 / 6 $ and the dimensionless run length $\ell_0 / D$~\cite{Omar2021pd}.

All simulations were performed using \texttt{HOOMD-Blue} and consisted of at least 55296 particles~\cite{Anderson2020}.
The primary purpose of our simulations was to inform the development of our bulk equations of state, ${\psi_V^*}$, ${p_C^{\rm bulk}}$, and ${\overline{U}}$ (or equivalently $p_{\rm act}^{\rm bulk}$), by measuring these properties in regions of the phase diagram where the \textit{system is spatially homogeneous}.
To determine these equations of state at high volume fractions (where a homogeneous solid is the stable configuration), simulations were initialized in a perfect fcc lattice (${\phi=\phi^{\rm CP}=0.74}$). 
The simulation box was periodically (and isotropically) expanded to reduce the volume fraction in increments of ${\Delta \phi = 0.0025}$.
At each volume fraction, the conservative and active contributions to the dynamic pressure along with the Steinhardt-Nelson-Ronchetti $q_{12}$ order parameter~\cite{Steinhardt1983} (taken to be $\psi_N^*$) were measured after the system was determined to have relaxed to a steady state.
Below an activity-dependent volume fraction, homogeneous states are no longer stable and a fluid nucleates.
This volume fraction can be quite high and, above an activity of ${\ell_0 / D \sim 1}$~\cite{Omar2021pd}, the \textit{only} observable stable solid phase is a nearly close-packed fcc crystal (see phase diagram in the main text), severely restricting the amount of data that can be obtained at high volume fractions.
Figure~\ref{sfig:PsSI} displays the contributions to the dynamic pressure obtained from this protocol. 

We also measure equations of state by initializing the system at a dilute volume fraction (${\phi = 0.05}$) and periodically compressing the simulation box (isotropically) to increase the volume fraction in increments of ${\Delta \phi = 0.025}$.
The locally stable configurations from this protocol corresponded to both globally stable and metastable fluids ($\psi_V^* \approx 0)$ with the measured pressures (not shown here) consistent with those of Ref.~\cite{Omar2023b}.
However, by determining the volume fraction at which these fluids develop a finite $\psi_V^*$, this protocol provides direct insight into the location of the order-disorder transition, $\phi^{\rm ODT}$.

Our simulations also allow us to extend the solid-fluid boundary reported in Ref.~\cite{Omar2021pd} to activities of $\ell_0/D < 0.9$.
These additional points are reported in the phase diagram displayed in the main text.

\subsection{\label{secSI:abpeos}Physical and Semi-Empirical Bulk Equations of State}

\begin{figure*}
	\centering
	\includegraphics[width=.99\textwidth]{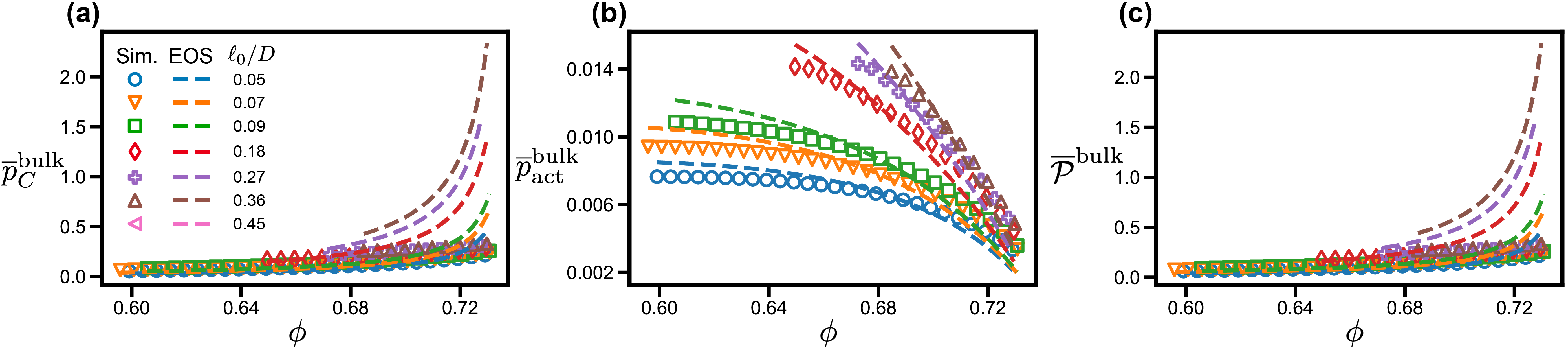}
	\caption{\protect\small{{Conservative interaction (${p^{\rm bulk}_C}$), active (${p_{\rm act}^{\rm bulk}}$), and total (${\mathcal{P}^{\rm bulk}}$) pressure equations of state at low activity. Pressures are nondimensionalized by the scale ${6 \zeta U_0 / \pi D^2}$.}}}
	\label{sfig:PsSI}
\end{figure*}

To construct the ABP solid-fluid phase diagram by applying our derived coexistence criteria, we need equations of state for the preferred crystallinity, ${\psi_N^*\left( \phi; \ell_0 / D \right)}$, and pressures, ${p^{\rm bulk}_C \left( \phi, \psi_N; \ell_0 / D \right)}$ and ${p_{\rm act}^{\rm bulk} \left( \phi, \psi_N; \ell_0 / D \right)}$, that accurately describe both fluid (${\psi_N \approx 0}$) and solid (${\psi_N > 0}$) phases at all activities.
We combine existing equations of state for an ABP fluid~\cite{Omar2023b} (developed for moderate activities ${\ell_0 / D > 1}$) and an equilibrium hard sphere fluid~\cite{Song1988} to develop accurate equations of state for ABP fluids at all activities.
To extend these equations of state to describe crystalline systems, we develop auxiliary equations of state [e.g.,~an equation of state for the maximum possible packing fraction, $\phi^{\rm max}(\psi_N; \ell_0/D)$] to capture the effects of nonzero $\psi_N$.

The active pressure of ABP fluids developed in Ref.~\cite{Omar2023b} $(p_{\rm act}^{\rm bulk})$ correctly recovers the ideal gas pressure in the reversible limit ($\ell_0 / D \rightarrow 0$), i.e.,~$p_{\rm act}^{\rm bulk} = \rho k_BT_{\rm act}$ where the active energy scale is ${k_BT_{\rm act} \equiv \zeta U_0\ell_0/6}$.
We extend $p_{\rm act}^{\rm bulk}$ to nonzero $\psi_N$ by introducing an equation of state $\phi^{\rm max} \left( \psi_N; \ell_0 / D \right)$ capturing the crystallinity-dependent maximum volume fraction:
\begin{subequations}
    \begin{align}
        \label{seq:pact}
        & p_{\rm act}^{\rm bulk} \left( \phi, \psi_N; \ell_0 / D \right) = \frac{\zeta U_0}{\pi D^2} \left( \frac{\ell_0}{D} \right) \phi \frac{1 + \tanh(A_{\rm act} \psi_N) \left[1 - c_{\rm act}^{(1)} t_{100}\left( \frac{\ell_0}{D}, \frac{\ell_0^c}{D} \right) + c_{\rm act}^{(2)} t_1 \left( \frac{\ell_0}{D}, \frac{\ell_0^{\rm act}}{D} \right) \right]}{1 + \bigg( 1 - \exp \left[-2^{7/6} \left( \frac{\ell_0}{D} \right) \right] \bigg) \frac{\phi}{1 - \phi / \phi^{\rm max} \left( \psi_N; \ell_0 / D \right)}}, \\
        & t_B \left( \ell_0 / D, \ell_0^* / D\right) = \frac{1 + \tanh \left(B \frac{\ell_0 - \ell_0^*}{D} \right)}{2}
    \end{align}
\end{subequations}
where $\ell_0^c = 18.8 \times 2^{-1/6} D$ is the MIPS critical point~\cite{Omar2021pd}, ${\phi^{\rm max} \left( \psi_N=0; \ell_0 / D \right) = \phi^{\rm RCP} = 0.645}$ to recover the fluid pressure in Ref.~\cite{Omar2023b}, and ${\phi^{\rm max} \left( \psi_N=1; \ell_0 / D \right) = \phi^{\rm CP} = 0.74}$ when the system has perfect crystalline order.
We fit the coefficients $A_{\rm act}=10$, $c_{\rm act}^{(1)}=29$, $c_{\rm act}^{(2)}=30$, and $\ell_0^{\rm act}=3 \times 2^{-1/6} D$.
The conservative interaction pressure in Ref.~\cite{Omar2023b} $(p_{C}^{\rm bulk, ABP})$ \textit{does not} recover the equilibrium hard sphere pressure $(p_{C}^{\rm bulk, HS})$~\cite{Song1988} in the low activity limit.
We remedy this by including an interpolation [through an equation of state $x \left( \ell_0 / D \right)$] between the conservative interaction pressures of an ABP fluid and an equilibrium hard sphere fluid (for which we use a simplified version of that in Ref.~\cite{Song1988} that we verify on Brownian dynamics simulations of passive hard spheres).
Extending $p_{C}^{\rm bulk, ABP}$ to nonzero $\psi_N$ requires an equation of state capturing an empirical crystallinity-induced slowing of its divergence [$\beta \left(\psi; \ell_0 / D \right)$] in addition to using $\phi^{\rm max} \left( \psi_N; \ell_0 / D \right)$ as the maximum volume fraction:
\begin{subequations}
    \label{seq:pc}
    \begin{align}
        \label{seq:pcinterp}
        & p^{\rm bulk}_C = x \big( \ell_0 / D \big) p^{\rm bulk, ABP}_C + \left[ 1 - x \left( \ell_0 / D \right) \right] p^{\rm bulk, HS}_C, \\
        \label{seq:pcABP}
        & p^{\rm bulk, ABP}_C \left( \phi, \psi_N; \ell_0 / D \right) = 2^{-7/6} \frac{\phi^2}{(1 - \phi / \phi^{\rm max})^\beta} \left(1 - c_C^{(1)} \tanh(A_C \psi_N) \left[1 - c_C^{(2)} t_{100}\left( \frac{\ell_0}{D}, \frac{\ell_0^c}{D} \right) \right] \right) \\
        \label{seq:pcHS}
        & p^{\rm bulk, HS}_C\left( \phi, \psi_N; k_B T \right) = \frac{6 k_B T}{\pi D^3} \frac{\phi}{1 - \phi / \phi^{\rm max}},
    \end{align}
\end{subequations}
where $\beta \left(\psi_N=0; \ell_0 / D \right)=1/2$ to recover the pressure in Ref.~\cite{Omar2023b} and $A_C=10$, $c_C^{(1)}=5/6$, and ${c_C^{(2)}=0.9}$ are coefficients we have fit.
We have introduced the thermal energy $k_B T$, which, in systems of active hard spheres, is generally density (and crystallinity) dependent and can be defined as $k_B T \equiv p_{\rm act}^{\rm bulk} / \rho$.
We find no appreciable differences in the resulting phase diagram when approximating this active temperature with that of ideal ABPs in 3D, ${k_BT = k_BT_{\rm act}=\zeta U_0 \ell_0 / 6}$~\cite{Takatori2014}, however.
We then use the simpler density-independent effective temperature, $k_BT_{\rm act}$, when constructing phase diagrams but note that the density dependence of the effective temperature may be more important for other systems.

The equations of state $x \left( \ell_0 / D \right)$, $\phi^{\rm max} \left( \psi_N; \ell_0 / D \right)$, and $\beta \left(\psi_N; \ell_0 / D \right)$ were empirically fit:
\begin{subequations}
    \label{seq:phimaxbeta}
    \begin{align}
        \label{seq:x}
        &x \big( \ell_0  / D \big) =  \left(1 - \tanh(A_x \psi_N)\right) \left[ \tanh\left(\log \left(\frac{\ell_0}{D} + 1 \right) + t_{10} \left( \frac{\ell_0}{D}, \frac{\ell_0^c}{D} \right) \left[e^{\ell_0 / D} - 1 \right] \right) \right] \nonumber \\& + \tanh(10 \psi_N) \left[ \tanh \left(\log\left(\frac{\ell_0}{D} + 1\right) + t_{100} \left( \frac{\ell_0}{D}, \frac{\ell_0^c}{D} \right)^{r_x^{(1)}} \left[e^{(\ell_0 / \ell_0^c)^{r_x^{(2)}}} - 1 \right] \right) \right], \\
        \label{seq:phimax}
        & \phi^{\rm max} \left( \psi_N ; \ell_0 / D \right) = \phi^{\rm RCP}  + \left( \phi^{\rm CP} - \phi^{\rm RCP} \right) \tanh \left( A_{\rm max} \psi_N \right), \\
        \label{seq:beta}
        & \beta \left( \psi_N; \ell_0 / D \right) =  \frac{1}{2} - \tanh(A_{\beta} \psi_N) \left[ c_{\beta}^{(1)} t_{5} \left( \ell_0 / D, \ell_0^c / D \right) + c_{\beta}^{(2)} t_{0.01} \left( \ell_0 / D, \ell_0^{\beta} / D \right) \right],
    \end{align}
\end{subequations}
where $A_x=10$, $r_x^{(1)}=r_x^{(2)}=10$, $A_{\rm max}=15.848$, $A_{\beta}=10$, $c_{\beta}^{(1)}=0.4$, $c_{\beta}^{(2)}=0.175$, and $\ell_0^{\beta}=50 \times 2^{-1/6} D$ are coefficients we have fit.
The forms of these fits were motivated by the previously discussed physical limits that we require to be met.

In order to use the equations of state in Eqs.~\eqref{seq:pact} and~\eqref{seq:pc} we require an equation of state for the preferred per-particle crystallinity, $\psi_N^*$, which we fit an expression for ${\psi_N^* \left( \phi; \ell_0 / D \right)}$ (see Fig.~1 in the main text):
\begin{multline}
    \label{seq:psistar}
    \psi_N^* \left( \phi; \ell_0 / D \right) = \Theta \left( \phi - \phi^{\rm ODT} \right) \tanh \bigg[ {\rm exp} \left( m_{\psi} \phi + c_{\psi} + A_{\psi} \frac{\phi}{\sqrt{1 - \phi / \phi^{\rm CP}}} \right) \\ \times \left( \Delta_{\psi}^{(1)} + \frac{ \left(\ell_0 / D\right)^{r_{\psi}^{(1)}}}{\Delta_{\psi}^{(2)} + \ln \left[\Delta_{\psi}^{(3)} + \left( \ell_0 / D \right)^{r_{\psi}^{(2)}} \right]} \right)^{-r_{\psi}^{(3)} \left( 1 - \phi / \phi^{\rm CP} \right)} \bigg],
\end{multline}
where ${m_{\psi}=18.8}$, ${c_{\psi}=-13.1}$, ${A_{\psi} = 0.05}$, ${\Delta_{\psi}^{(1)}=0.01}$, ${\Delta_{\psi}^{(2)}=\Delta_{\psi}^{(3)}=1}$, $r_{\psi}^{(1)}=1.16$ and ${r_{\psi}^{(2)}=r_{\psi}^{(3)}=2}$, are again constants that have been fit.
The equation of state for the order-disorder volume fraction, ${\phi^{\rm ODT} \left( \ell_0 / D \right)}$, [see the inset of Fig.~1 in the main text] was determined to be:
\begin{equation}
    \label{seq:phiodt}
    \phi^{\rm ODT} \left(\ell_0 / D \right) = \phi^{\rm ODT}_{\rm eqm} + \left( \phi^{\rm RCP} - \phi^{\rm ODT}_{\rm eqm} \right) \tanh \left(A_{\rm ODT} \log\left[c_{\rm ODT} \frac{\ell_0}{D} + 1\right] \right),
\end{equation}
where $\phi^{\rm ODT}_{\rm eqm} = 0.515$ is the equilibrium hard sphere $\phi^{\rm ODT}$ and $A_{\rm ODT} = 1.381$ and ${c_{\rm ODT}=0.909}$ are fitted constants.

We see that since our equation for $\psi_N^*$ in Eq.~\eqref{seq:psistar} experiences a discontinuity at $\phi^{\rm ODT}$, our equation for $p^{\rm bulk}_C$ in Eq.~\eqref{seq:pc} does as well.
This discontinuity is necessary for passive solid-fluid coexistence, as the pressure (evaluated at $\psi_N^*$) must be non-monotonic with increasing $\rho$ in order to find binodals.

Figure ~\ref{sfig:PsSI} shows the fits for ${p^{\rm bulk}_C}$ and ${p_{\rm act}^{\rm bulk}}$ at low activities after inserting the expressions for $x$, ${\phi^{\rm max}}$, ${\beta}$, $\phi^{\rm ODT}$, and $\psi^*$ into Eqs.~\eqref{seq:pact} and~\eqref{seq:pc}.
While the fit for ${p^{\rm bulk}_C}$ is an overestimate, the qualitative ${\ell_0  /D}$ and ${\phi}$ dependent trends are captured, whereas the fit for $p_{\rm act}^{\rm bulk}$ is more quantitatively accurate.

\subsection{Characterization of the ``Pseudo''-spinodal}

\begin{figure*}
	\centering
	\includegraphics[width=.95\textwidth]{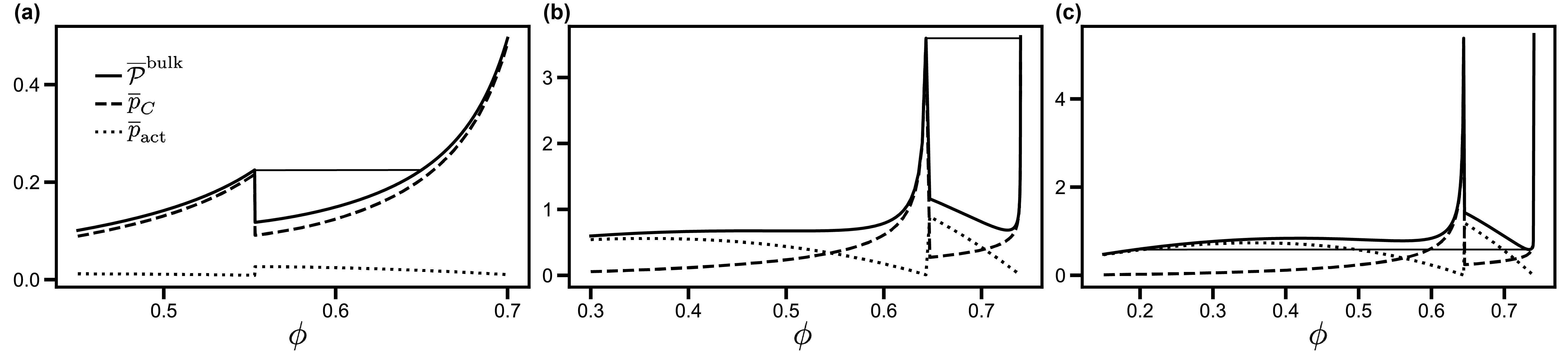}
	\caption{\protect\small{{ABP pressure (nondimensionalized by the scale ${6 \zeta U_0 / \pi D^2}$) at (a) low activity ${\ell_0 / D = 0.9}$, (b) intermediate activity ${\ell_0 / D = 17.4}$, and (c) high activity ${\ell_0 / D = 22.3}$. ${p^{\rm bulk}_C}$ is shown in dashed lines while ${p_{\rm act}^{\rm bulk}}$ is shown in dotted lines.}}}
	\label{sfig:PsSI2}
\end{figure*}

There are two spinodals, or regions of instability, in our dynamic pressure $\left( \mathcal{P}^{\rm bulk} = p_C^{\rm bulk} + p_{\rm act}^{\rm bulk} \right)$ of active hard spheres in Eqs.~\eqref{seq:pc} and \eqref{seq:pact}.
The first is a true spinodal indicating that the fluid phase ($\psi_N \approx 0$) is unstable at certain densities.
The fluid spinodal, which occurs above the critical activity, arises from a non-monotonic active pressure and results in MIPS.
The second is a ``pseudo''-spinodal which drives crystallization, even in the reversible limit.
We distinguish this spinodal as it indicates that states of intermediate density and finite $\psi_N$ (which cannot generally be prepared) are unstable. 

For a solid-fluid transition to occur for passive hard spheres, ${p^{\rm bulk}_C}$ must contain a discontinuity at the order-disorder volume fraction, $\phi^{\rm ODT}$.
This discontinuity represents a region of instability that occurs over an infinitely narrow range of ${\phi}$ where $\psi_N^*$ adopts a nonzero value, representing a pseudo-spinodal.
The pseudo-spinodal widens at finite activity due to the non-monotonicity of ${p_{\rm act}^{\rm bulk}}$, encompassing a finite range of volume fractions above $\phi^{\rm ODT}$.
Figure~\ref{sfig:PsSI2} shows the widening of this pseudo-spinodal, showing the active and conservative interaction contributions to ${\mathcal{P}^{\rm bulk}}$ at low, intermediate, and high activity (the same activities as Fig.~2 in the main text).